\documentclass[twocolumn]{aastex631}

\usepackage{xcolor,longtable}

\begin{document}

\title{The Science Performance of the Gemini High Resolution Optical Spectrograph}

\author[0000-0003-4666-6564]{Alan W. McConnachie}
\affiliation{National Research Council Herzberg Astronomy and Astrophysics, 5071 West Saanich Road, Victoria, B.C., V8Z6M7, Canada}
\affiliation{Department of Physics and Astronomy, University of Victoria, Victoria, BC V8W 3P2, Canada}
\author[0000-0003-2969-2445]{Christian R. Hayes}
\affiliation{National Research Council Herzberg Astronomy and Astrophysics, 5071 West Saanich Road, Victoria, B.C., V8Z6M7, Canada}
\author[0000-0001-5528-7801]{J. Gordon Robertson}
\affiliation{Australian Astronomical Optics, Macquarie University, 105 Delhi Rd, North Ryde NSW 2113, Australia}
\affiliation{Sydney Institute for Astronomy, School of Physics, The University of Sydney, NSW 2006, Australia}
\author{John Pazder}
\affiliation{National Research Council Herzberg Astronomy and Astrophysics, 5071 West Saanich Road, Victoria, B.C., V8Z6M7, Canada}
\author[0000-0002-6194-043X]{Michael Ireland}
\affiliation{Research School of Astronomy and Astrophysics, College of Science, Australian National University, Canberra 2611, Australia} 
\author{Greg Burley}
\affiliation{National Research Council Herzberg Astronomy and Astrophysics, 5071 West Saanich Road, Victoria, B.C., V8Z6M7, Canada}
\author{Vladimir Churilov}
\affiliation{Australian Astronomical Optics, Macquarie University, 105 Delhi Rd, North Ryde NSW 2113, Australia}
\author{Jordan Lothrop}
\affiliation{National Research Council Herzberg Astronomy and Astrophysics, 5071 West Saanich Road, Victoria, B.C., V8Z6M7, Canada}
\author{Ross Zhelem}
\affiliation{Australian Astronomical Optics, Macquarie University, 105 Delhi Rd, North Ryde NSW 2113, Australia}
\author[0000-0002-4641-2532]{Venu Kalari}
\affiliation{Gemini Observatory/NSF’s NOIRLab, Casilla 603, La Serena, Chile}
\author{Andr{\'e} Anthony}
\affiliation{National Research Council Herzberg Astronomy and Astrophysics, 5071 West Saanich Road, Victoria, B.C., V8Z6M7, Canada}
\author{Gabriella Baker}
\affiliation{Australian Astronomical Optics, Macquarie University, 105 Delhi Rd, North Ryde NSW 2113, Australia}
\author[0000-0002-2606-5078]{Trystyn Berg}
\affiliation{European Southern Observatory, Alonso de Cordova, 3107, Casilla, 19001 Santiago, Chile}
\author{Edward L. Chapin}
\affiliation{National Research Council Herzberg Astronomy and Astrophysics, 5071 West Saanich Road, Victoria, B.C., V8Z6M7, Canada}
\author{Timothy Chin}
\affiliation{Australian Astronomical Optics, Macquarie University, 105 Delhi Rd, North Ryde NSW 2113, Australia}
\author{Adam Densmore}
\affiliation{National Research Council Herzberg Astronomy and Astrophysics, 5071 West Saanich Road, Victoria, B.C., V8Z6M7, Canada}
\author{Ruben Diaz}
\affiliation{Gemini Observatory/NSF’s NOIRLab, Casilla 603, La Serena, Chile} 
\author{Jennifer Dunn}
\affiliation{National Research Council Herzberg Astronomy and Astrophysics, 5071 West Saanich Road, Victoria, B.C., V8Z6M7, Canada}
\author{Michael L. Edgar}
\affiliation{Australian Astronomical Observatory}
\author{Tony Farrell}
\affiliation{Australian Astronomical Optics, Macquarie University, 105 Delhi Rd, North Ryde NSW 2113, Australia}
\author[0000-0002-4628-3726]{Veronica Firpo}
\affiliation{Gemini Observatory/NSF’s NOIRLab, Casilla 603, La Serena, Chile} 
\author{Javier Fuentes}
\affiliation{Gemini Observatory/NSF’s NOIRLab, Casilla 603, La Serena, Chile}
\affiliation{European Southern Observatory, Alonso de C{\'o}rdova 3107, Vitacura, Regi{\'o}n Metropolitana Santiago de Chile, Chile}
\author{Manuel Gomez-Jimenez}
\affiliation{Gemini Observatory/NSF’s NOIRLab, Casilla 603, La Serena, Chile} 
\author{Tim Hardy}
\affiliation{National Research Council Herzberg Astronomy and Astrophysics, 5071 West Saanich Road, Victoria, B.C., V8Z6M7, Canada}
\author{David Henderson}
\affiliation{Gemini Observatory/NSF’s NOIRLab, 670 N. A’ohoku Place, Hilo, Hawai’i 96720 USA}
\author{Alexis Hill}
\affiliation{National Research Council Herzberg Astronomy and Astrophysics, 5071 West Saanich Road, Victoria, B.C., V8Z6M7, Canada}
\author[0000-0002-6633-7891]{Kathleen Labrie}
\affiliation{Gemini Observatory/NSF’s NOIRLab, 670 N. A’ohoku Place, Hilo, Hawai’i 96720 USA}
\author[0000-0002-4350-7632]{Jaclyn Jensen}
\affiliation{Department of Physics and Astronomy, University of Victoria, Victoria, BC V8W 3P2, Canada}
\author{Sam Lambert}
\affiliation{National Research Council Herzberg Astronomy and Astrophysics, 5071 West Saanich Road, Victoria, B.C., V8Z6M7, Canada}
\author{Jon Lawrence}
\affiliation{Australian Astronomical Optics, Macquarie University, 105 Delhi Rd, North Ryde NSW 2113, Australia}
\author{G. Scott Macdonald}
\affiliation{National Research Council Herzberg Astronomy and Astrophysics, 5071 West Saanich Road, Victoria, B.C., V8Z6M7, Canada}
\author{Steven Margheim}
\affiliation{Rubin Observatory/NSF’s NOIRLab, Casilla 603, La Serena, Chile}
\author{Bryan Millar}
\affiliation{Gemini Observatory/NSF’s NOIRLab, Casilla 603, La Serena, Chile} 
\author{Rolf Muller}
\affiliation{Australian Astronomical Optics, Macquarie University, 105 Delhi Rd, North Ryde NSW 2113, Australia}
\author[0000-0003-4685-4231]{Jon G. Nielsen}
\affiliation{Research School of Astronomy and Astrophysics, College of Science, Australian National University, Canberra 2611, Australia}
\author{Gabriel P{\'e}rez}
\affiliation{Gemini Observatory/NSF’s NOIRLab, Casilla 603, La Serena, Chile} 
\author[0000-0001-5558-6297]{Carlos Quiroz}
\affiliation{Gemini Observatory/NSF’s NOIRLab, Casilla 603, La Serena, Chile} 
\author[0000-0001-7518-1393]{Roque Ruiz-Carmona}
\affiliation{Gemini Observatory/NSF’s NOIRLab, Casilla 603, La Serena, Chile} 
\author{Kim M. Sebo}
\affiliation{Research School of Astronomy and Astrophysics, College of Science, Australian National University, Canberra 2611, Australia} 
\author[0000-0002-3182-3574] {Federico Sestito}
\affiliation{Department of Physics and Astronomy, University of Victoria, Victoria, BC V8W 3P2, Canada}
\author{Kareleyne Silva}
\affiliation{Gemini Observatory/NSF’s NOIRLab, Casilla 603, La Serena, Chile}
\author{Chris Simpson}
\affiliation{Gemini Observatory/NSF’s NOIRLab, 670 N. A’ohoku Place, Hilo, Hawai’i 96720 USA}
\author{Greg Smith}
\affiliation{Australian Astronomical Optics, Macquarie University, 105 Delhi Rd, North Ryde NSW 2113, Australia}
\author{Sudharshan Venkatesan}
\affiliation{Australian Astronomical Optics, Macquarie University, 105 Delhi Rd, North Ryde NSW 2113, Australia}
\author{Fletcher Waller}
\affiliation{Department of Physics and Astronomy, University of Victoria, Victoria, BC V8W 3P2, Canada}
\author{Lewis Waller}
\affiliation{Australian Astronomical Optics, Macquarie University, 105 Delhi Rd, North Ryde NSW 2113, Australia}
\author{Ivan Wevers}
\affiliation{National Research Council Herzberg Astronomy and Astrophysics, 5071 West Saanich Road, Victoria, B.C., V8Z6M7, Canada}
\author[0000-0003-4134-2042] {Kim A. Venn}
\affiliation{Department of Physics and Astronomy, University of Victoria, Victoria, BC V8W 3P2, Canada}
\author[0000-0002-2565-1964]{Peter Young}
\affiliation{Research School of Astronomy and Astrophysics, College of Science, Australian National University, Canberra 2611, Australia} 

\begin{abstract}

The Gemini High Resolution Optical Spectrograph (GHOST) is a fiber-fed spectrograph system on the Gemini South telescope that provides simultaneous wavelength coverage from 348 -- 1061\,nm, and designed for optimal performance between 363 -- 950\,nm. It can observe up to two objects simultaneously in a 7.5\,arcmin diameter field of regard at $R \simeq 56,000$ or a single object at $R \simeq 75,000$. The spectral resolution modes are obtained by using integral field units to image slice a 1.2” aperture by a factor of five in width using 19 fibers in the high resolution mode and by a factor of three in width using 7 fibers in the standard resolution mode. GHOST is equipped with hardware to allow for precision radial velocity measurements, expected to approach meters per second precision. Here, we describe the basic design and operational capabilities of GHOST, and proceed to derive and quantify the key aspects of its on-sky performance that are of most relevance to its science users.

\end{abstract}


\section{Introduction}
\label{sect:intro}  


The canon of scientific discoveries that have been enabled by high resolution spectroscopy, especially in the era of $8 - 10$\,m-class facilities, is extensive. No single paragraph can do an adequate summary to either the instruments or the science that is contained in this very broad field. Inadequate summaries, on the other hand, are certainly possible. For example, using the Very Large Telescope's (VLT's) Ultraviolet and Visual {\'E}chelle Spectroph (UVES; \citealt{dekker2000}), the first measurement of the temperature of the Cosmic Microwave Background at high redshift was made by observing Carbon Monoxide in high redshift quasars (\citealt{srianand2008, noterdaeme2011}). This same instrument, working in conjuction with the High Resolution {\'E}chelle Spectrometer (HIRES; \citealt{vogt1994}) at the W.M. Keck Observatory, contributed to the first precision measurement of the primordial abundance of deuterium by examination of a metal poor, damped Lyman-$\alpha$ system at $z \sim 3$. The Magellan Inamori Kyocera {\'E}chelle (MIKE) Spectrograph (\citealt{bernstein2003}) has shown that at least one of the dwarf galaxy companions to the Milky Way is an excellent laboratory for the study of the rare events that are the nucleosynthetic origin of the heaviest elements in the periodic table through the elusive r-process (\citealt{ji2016a, ji2016b}). More generally, the High Dispersion Spectrograph (HDS; \citealt{noguchi2002}) on Subaru, and indeed all of the major high resolution instruments, have contributed significantly to better understanding chemical nucleosynthesis through study of some of the oldest and/or most metal-poor stars in our galaxy and its neighbours (e.g., \citealt{honda2004, aoki2005, aoki2006}). Where major multiplexing capabilities are available, for example with the VLT Fiber Large Array Multi-Element Spectrograph (FLAMES; \citealt{pasquini2002}), the Multi-Mirror Telescope's Hectochelle (\citealt{szentgyorgyi2011}), or the Michigan/Magellan Fibre System (M2FS; \citealt{mateo2012}), high resolution spectrographs have contributed tosome of the largest and most significant datasets concerning the internal dynamics of Milky Way dwarf galaxies and the dark matter halos in which they are expected to reside (e.g., \citealt{tolstoy2006, battaglia2006, battaglia2008, battaglia2011, walker2007a, walker2007b, walker2009b, walker2015b}). Finally, while the original and field-defining discoveries of exoplanets using the Doppler effect were made using spectrographs on smaller telescopes, larger facilities have continued to play an essential role in the ongoing discovery of these systems, their characterisation and the characterisation of the host star populations (e.g., \citealt{santos2004, brewer2016, petigura2017}),

It is against this impressive scientific backdrop that the Gemini High Resolution Optical Spectrograph (GHOST) makes its debut. GHOST is a newcomer to a mature field, and its design and functionality has been inspired by the successes of its predecessors. The goal is for GHOST to be a go-to instrument for a broad swath of research by leveraging the latest in technological advancements to increase its sensitivity and general scientific utility. The purpose of this contribution is to describe its relevant design and operational features, and to derive and quantify its scientific performance from the perspective of the science user using on-sky (commissioning) data, with the intent of providing the international community with some insights into how GHOST will be of utility for their science and discoveries yet to be made.

GHOST has been designed, built and commissioned by a collaboration involving the AAO-Macquarie, the National Research Council Herzberg Astronomy and Astrophysics Research Center (NRC-HAA) and the Australian National University working with the Gemini Observatory. AAO-Macquarie are project leads and responsible for the Cassegrain Unit, fiber feed system and slit viewer; NRC-HAA designed and built the bench spectrograph unit, and ANU were responsible for instrument software and the data reduction system (DRS). 

Kick-off for the Preliminary Design of GHOST was in May 2014. The Cassegrain Unit and Fiber System was commissioned independently of the bench spectrograph in 2018. The bench spectrograph completed integration and testing at NRC-HAA in late 2019 and was shipped to Gemini South in February 2020.  The following months saw the realisation of an unanticipated risk register item with global repercussions, that led to a two year delay. Then, in March 2022, the team was able to reach Gemini South to perform integration and testing. 

Science Commissioning of GHOST was initially scheduled for nine nights, from June 20 - 28 2022, and which saw the full Commissioning Team led by NRC-HAA on-site at Gemini South. The first part of the run had to content with extremely challenging weather conditions which resulted in extending the run by 1.5 nights, and conditions during nights 9, 10 and 11 were much better and resulted in good data that allowed for a range of successful on-sky testing to occur. A second commissioning run was scheduled for five nights, September 12 - 16 2022, where the external team connected remotely and Gemini staff led the campaign. Brief overviews of these runs can be found in \cite{mcconnachie2022a} and \cite{mcconnachie2022b}. Gemini staff also obtained subsequent on-sky data with GHOST on the nights of December 8 2022, January 28 - 29 2023 and February 13 - 14 2023. All the on-sky data that forms the basis of the analyses presented in this paper were obtained either during the commissioning runs or on these nights. Gemini has continued to schedule on-sky time for GHOST since February 2023, although we do not use any of these data in what follows.

The focus of the ensuing discussion and analysis is primarily on the delivered performance of GHOST and what it means from a science user perspective. A brief overview of the design will of course be given, but readers wanting more design information should refer to \cite{ireland2012, ireland2014} for details on the
overall instrument system; \cite{zhelem2018, zhelem2020} and \cite{churilov2018} for the Cassegrain unit and fiber; \cite{pazder2016} for the optical design of the
bench spectrograph; \cite{pazder2022a} for the design of the lens barrels; \cite{lothrop2020} for the spectrograph enclosure; \cite{young2016} and \cite{nielsen2018} for the instrument control software; \cite{ireland2016, ireland2018} and \cite{hayes2022}
for discussion of the precision radial velocity mode and data reduction pipeline; \cite{macdonald2022} for details on the methodologies employed during shipping of the bench spectrograph. The opto-mechanical design of the bench spectrograph will be the subject of a future contribution (Anthony et al., {\it in preparation}). An accompanying paper will describe the original science drivers of GHOST, its data reduction pipeline, and its integration into Gemini operations (including any minor changes the latter might make to its offered capabilities compared to that described here; Kalari et al., {\it in preparation}). 

This paper is organised as follows. Section 2 provides an overall instrument design description, focusing on those aspects that make GHOST especially notable relative to its peers. Section 3 describes the key operational characteristics of GHOST that are relevant to its science users. Section 4 presents the on-sky performance of GHOST in relation to its key science-enabling properties. Section 5 provides a summary. 

\section{GHOST Instrument Design}

\subsection{Overview}

GHOST was initially conceived by the Gemini Observatories to be a workhorse instrument expected to meet a wide range of science goals. Gemini originally provided 14 science cases derived from earlier community White Papers. The majority of these involved measuring chemical abundances in different astrophysical contexts. These science cases were augmented by the GHOST science team to include precision velocity science cases and quasar absorption lines studies. From this suite of science cases, the driving science requirements were identified, including complete spectral coverage across the optical band at high sensitivity, operation at moderate and high spectral resolutions, and the ability to provide velocity precision approaching meters per second in the highest resolution setting. Details of the original science drivers and requirement flowdown can be found in \cite{ireland2012, ireland2014}, and will be discussed in more detail in  Kalari et al., {\it in preparation}.

Table~\ref{tab:summary} provides an overview of the key GHOST instrument parameters. GHOST is  a fiber-fed spectrograph system with simultaneous wavelength coverage from  $348 - 1061$\,nm, and designed for optimal performance from $363 - 950$\,nm. It can observe up to two objects simultaneously in a 7.5\,arcmin diameter field of regard at $R = \lambda / \Delta \lambda \simeq 56,000$ or a single object at $R \simeq 75,000$. GHOST consists of three primary components; the Cassegrain unit mounted on the telescope, the spectrograph bench located in the pier lab, and a fiber cable connecting the two. The Cassegrain unit contains the positioning arm system for two micro-lens based integral field unit (IFU) systems. The IFUs image slice a 1.2” aperture by a factor of three in width using 7 fibers in the standard resolution mode and by a factor of five in width using 19 fibers in the high resolution mode. The ability to observe two objects simultaneously at standard resolution allows for an increase in observing efficiency for some science programs, and the length of the two slits combined closely matches that of the slit for the single object high resolution mode, while fitting on the detector with adequate sampling.  A 32m fiber cable transports the light to the bench spectrograph that is located in a thermally stabalized enclosure in the Gemini Pier lab, and the fibers reformat the sliced image into a slit for injection into the spectrograph. Acquisition and guide fibers feed a guide camera for fine centering of the IFUs on the science targets at acquisition and during the exposure. A slit unit camera provides for active monitoring of the slit illumination during an exposure.

\begin{figure*}
   \begin{center}
   \includegraphics[width=16cm]{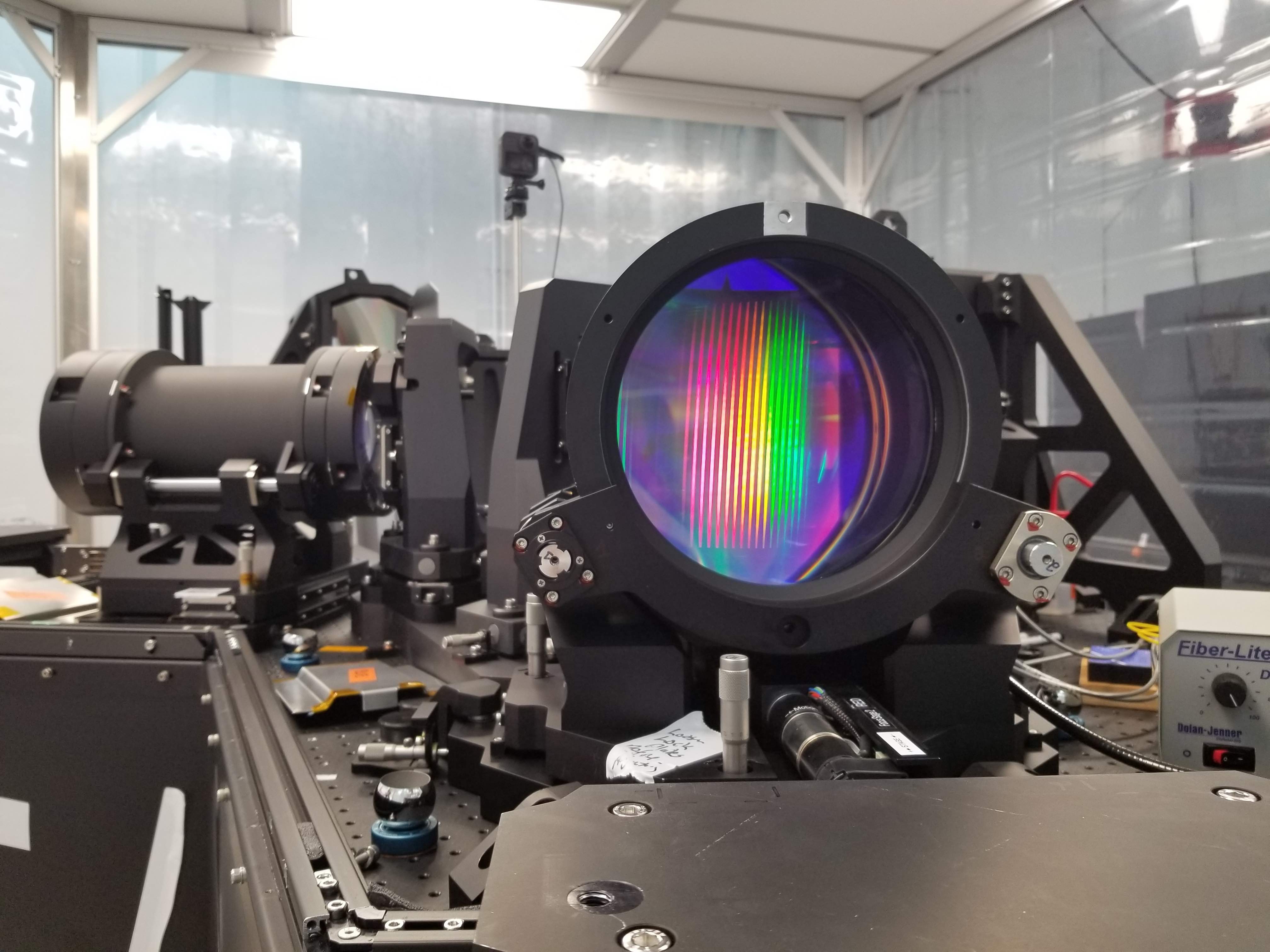}
   \end{center}
   \caption{\label{photo} A view of GHOST during integration at Gemini South, taken from the location of the red detector looking into the red camera lens barrel as a white light source is injected into the system.}
\end{figure*} 

\begin{figure*}
   \begin{center}
   \includegraphics[width=10cm]{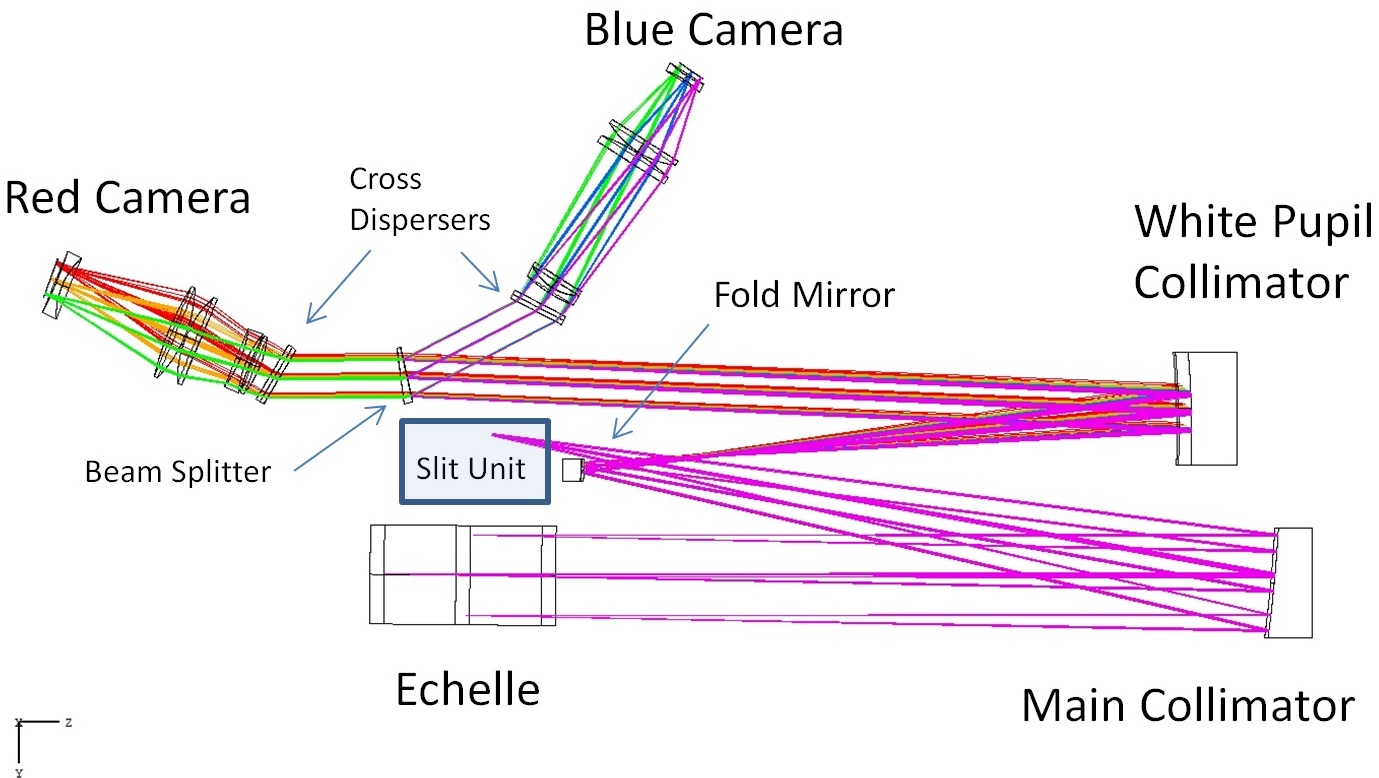}
   \end{center}
   \caption{\label{optical_design} The optical layout of the GHOST bench spectrograph.}
\end{figure*}

The bench spectrograph is a two arm, R2 {\'e}chelle white-pupil design, using Volume Phase Holographic (VPH) gratings for cross-dispersion. The white pupil relay is a zero-Petzval sum design to eliminate field curvature at the cross disperser grating. The cameras are designed with linear colour and tilted detectors. The longitudinal color in the camera is corrected to match the tilted detector, thus the color aberration in the lens varies in a linear way from the bottom to the top of the echellegram. In addition, with tilted detectors, the reflections from the detector are not directed back into the system. This eliminates the need for exotic glasses and minimizes ghosts. The beam is split by a high efficiency beam splitter which feeds the red and blue VPH gratings and the respective camera optics and detectors. GHOST is designed so that the main science orders in the blue are $m = 95 - 64$, and the main science orders in the red are $m = 66 - 34$. In practice, however, additional orders are incident on the detectors which allows for expanded spectral coverage. Figure~\ref{photo} shows a view of GHOST during integration at Gemini South, taken from the location of the red detector system looking into the camera as a white light source is injected into the system. The many orders of the red channel are clearly visible. An overview of the optical design of the bench spectrograph is shown in Figure~\ref{optical_design}.

The thermal enclosure is designed as a box in a box. The outer box has active heaters and sensors embedded in the insulated walls. These embedded heaters maintain a $20^\circ$C temperature on all interior surfaces of the outer shell. An inner shell provides a second layer of thermal isolation for the spectrograph. Heat sources are minimized within the inner shell. Additionally, there is a water cooling system to remove energy from any heat sources within the enclosure. The spectrograph bench and optical mounts are aluminum to minimize thermal distortions with temperature changes. The optical mechanical structure is designed with universal hold downs and micro-adjustment flexures for alignment.

GHOST uses the Gemini Calibration System (GCAL; \citealt{ramsay2000}) to obtain the usual calibration exposures. Briefly, GCAL is mounted on the Instrument Support Structure (ISS) of Gemini, and light can be directed from GCAL into any of the ISS-mounted instruments. GCAL mimics the f/16 beam of the telescope, and so provides sources suitable for calibration to the instruments. For GHOST, a new 100W halogen flat-field lamp was installed and is used to obtain flat-field frames at high SNR in short exposures. In addition, a new ThAr source was installed and is used for wavelength calibration of GHOST. GHOST has a ThXe internal calibration lamp, that can be used simultaneously with a science exposure at high resolution, with the light fed down a dedicated calibration fiber onto the detector adjacent to the science data. Fiber agitators are available to reduce modal noise when necessary for very high signal-to-noise ratio (SNR) observations. The choice was made to use ThXe as the internal calibration lamp because, while ThAr is the more common calibration source for spectrographs shortward of $\sim 7000\AA$, very strong lines of Argon tend to bleed and make many CCD pixels unusable at red wavelengths. Xenon, however, does not encounter this problem until even redder wavelengths. Further, the higher atomic mass of Xenon means that, in principle, its lines are narrower and therefore more suited to PRV applications.

\begin{table*}
\begin{center}
\begin{tabular*}{0.95\textwidth}{l|l}
  Wavelength range & (348) 363 – 950 (1061) nm\\
  \hline
  {\'E}chelle Grating&Richardson Grating Labs 424E, $65^\circ$ blaze (R 2.14)\\
  & $\gamma = 0.56^\circ$\\
&$G= 52.67$ grooves/mm\\
                   &Ruled area: 204mm x 410mm\\
  \hline
  Collimator & 1750.6 mm focal length, at f/10 \\
  & 176 mm beam diameter at {\'e}chelle\\
\hline
White pupil magnification &0.45\\
\hline
Beam Splitter & Cut on wavelength: 529.4nm (center of order 65) \\
&  Order Overlap: $\pm 1.5$ orders\\&           Reflection Band: $R > 97\%$ from 360nm to 517.4 nm\\
                   &         Transmission band: $T > 96\%$ from 541.4nm to 1000nm\\
  & Substrate: N-BK7HT 210mm $\times$ 126mm $\times$ 15mm, $5^\circ$ wedge\\
  \hline
Cross dispersers& Blue: 1050 lines/mm VPHG, $a = 35^\circ$\\
&  Substrate: Fused Silica 129mm $\times$ 140mm  $\times$ 10mm  $\times$ 2 \\
  &Red: 500 lines/mm VPHG, $a = 31^\circ$\\
                     &Substrate: Fused Silica 129mm  $\times$ 140mm  $\times$ 10mm  $\times$ 2 \\
 & Beam Diameter of recorded VPHG: 127mm\\
  \hline
Blue Camera &Orders: (98) 95 to 64\\
&Minimum order width: $1.12 \times$  FSR\\
&Wavelength range: (348) 363 to 544nm\\
&Field of view: $14.3^\circ$ (full angle)\\
&Focal length: 295mm\\
&CCD: e2v ccd231-84 4k$\times$4k pixel detector area, 15$\mu$ m pixels\\
  \hline
Red Camera& Orders: 66 to 34 (32) \\
&Minimum order width: $1.12 \times$ FSR\\
&Wavelength range: 521 to 950 (1061) nm\\
&Field of view: $25.7^\circ$(full angle)\\
&Focal length: 295mm\\
&CCD: e2v ccd231-c6 6kx6k pixel detector area, 15$\mu$ m pixels\\
\end{tabular*}
\caption{\label{tab:summary}High level design summary of the main components of the GHOST spectrograph. Orders and wavelengths in parentheses correspond to the orders and wavelengths that fall on the detector, whereas the values outside of parentheses correspond to the values for which the design is optimised. The detectors are summarised in more detail in Table~\ref{tab:detectors}. }
\end{center}
\end{table*}

\subsection{Design highlights}

\label{sec:design}

Here, we highlight a few design features of GHOST that are notable relative to its peers and which positively impact the quality of science data that is obtained by the user.






\subsubsection{Microlenses and image slicing}

\label{sec:mapping}

\begin{figure*}
   \begin{center}
   \includegraphics[width=14cm]{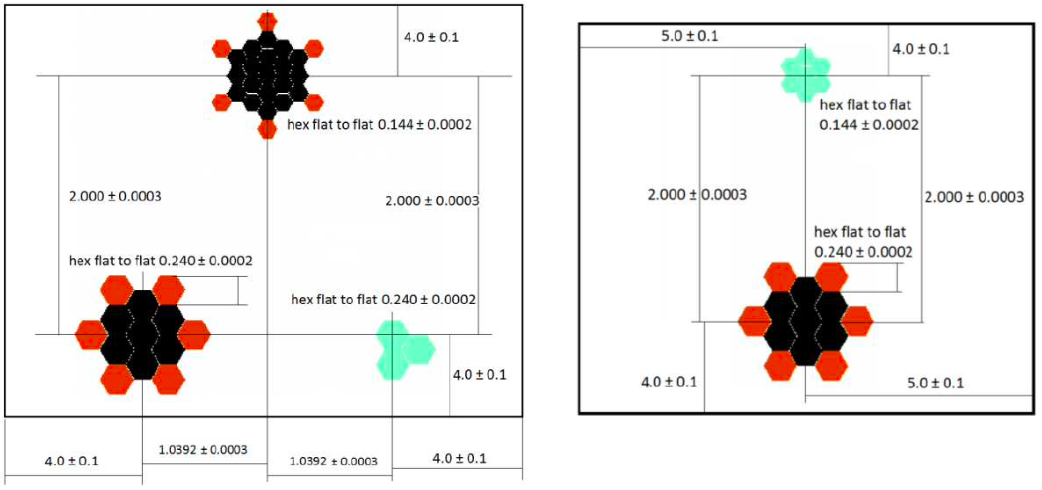}
   \end{center}
   \caption{ \label{microlens_in} Arrangement of hexagonal microlenses in the image plane for IFU1 (left panel) and IFU2 (right panel). All measurements are in millimeters; the image scale is 0.61mm/arcsec. Black hexagons correspond to the main science (target) regions; green hexagons correspond to sky regions, and orange hexagons correspond to acquisition and guiding fibers. Large hexagons correspond to the standard resolution mode, and small hexagons correspond to the high resolution mode. For clarity, these figures are not to scale.}
\end{figure*} 

Light from an astronomical target is injected into the fiber cables with the help of microlenses. Specifically, each fiber feed consists of an array of microlens doublets that provide a telecentric reimaging of the entrance pupil on the fiber core. The first of these microlenses are hexagonal with the configuration shown in Figure~\ref{microlens_in}. The overall sizes match the typical seeing disc at Gemini, and they have individual sizes corresponding to 0.4 or 0.25 arcsec on sky, for standard and high resolution modes, respectively. These segment the star image and set the spatial sampling of the fiber input. The second, circular, lens array collimates the flux from different field positions, and images the telescope pupil on to the fiber core in parallel light. The pupil diameter is 50$\mu$m for both resolution modes, and the optical prescription for the injection optics is the same for both modes, except for the aperture size. As a result, the standard-resolution mode propagates through the fiber at f/2.8, whereas the high-resolution mode is proportionally slower at f/4.7. 

Figure~\ref{microlens_in} shows the image-plane microlens configurations for each of the ``IFU heads'' (IFU head 1 in the left panel, hereafter IFU1, IFU head 2 in the right panel, hereafter IFU2), where all measurements are in millimeters. The image scale is 0.61mm/arcsec. Each IFU head consists of three (IFU1) and two (IFU2) individual IFUs with a fixed configuration relative to each other. Everything in the left panel (IFU1) moves together on a single positioner, and everying in the right panel (IFU2) moves together on a single positioner. Black microlenses correspond to the main science regions. Each of the two IFU heads has a standard resolution mode (larger hexagons) and a high resolution mode (smaller hexagons). In addition, green hexagons correspond to sky fibers, that are also routed to the spectrograph. Thus it is possible to observe two science targets and sky simultaneously in standard resolution, or one target and sky simultaneously in high resolution; see Section~\ref{sec:fp} for more details. Also visible are acquisition and guiding fibers in orange, that are routed to the acquisition and guiding camera, discussed in Section~\ref{sec:ag}. 

\begin{figure}
   \begin{center}
   \includegraphics[width=8cm]{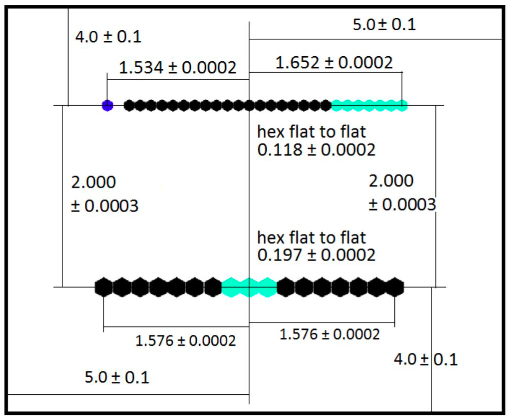}
   \end{center}
   \caption{ \label{microlens_out} Arrangement of microlenses at the output of the fibers/input to the {\'e}chelle spectrograph. The high resolution microlenses are at the top, the standard resolution microlenses are at the bottom. Color coding and sizes are the same as in Figure~\ref{microlens_in}. Also shown as the blue hexagon is the location of the microlens corresponding to the internal calibration source available in high resolution mode. For clarity, this figure is not to scale.}
\end{figure} 

At the output of the fibers (input to the {\'e}chelle spectrograph), the fibers are reformatted into a linear slit. Two microlenses per fiber form a pseudo-slit which is dispersed by the spectrograph. Whereas at the input end, one lens performs the function of an image slicer and the other one ensures telecentric injection for optimum guidance through a fiber, at the slit end the lenses work in reverse. The microlens configuration at the spectrograph slit is shown in in Figure~\ref{microlens_out}, where the high resolution slit is at the top and the standard resolution slit is at the bottom. Colors correspond to Figure~\ref{microlens_in} and all measurements are again in millimeters. Also shown as a dark blue hexagon is the internal calibration fiber that can operate simultaneously with the high resolution observations, and which will be discussed in Section~\ref{sec:prv}.

\begin{figure}
   \begin{center}
   \includegraphics[width=8cm]{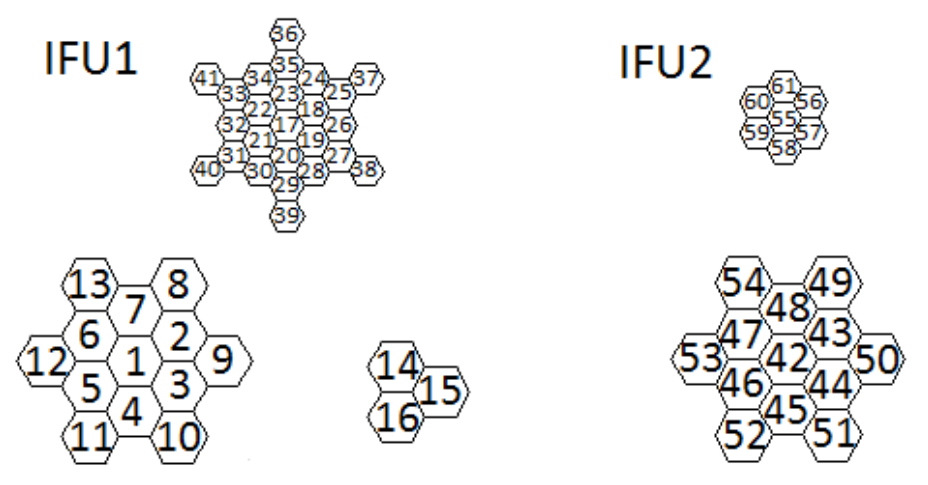}
   \includegraphics[width=8cm]{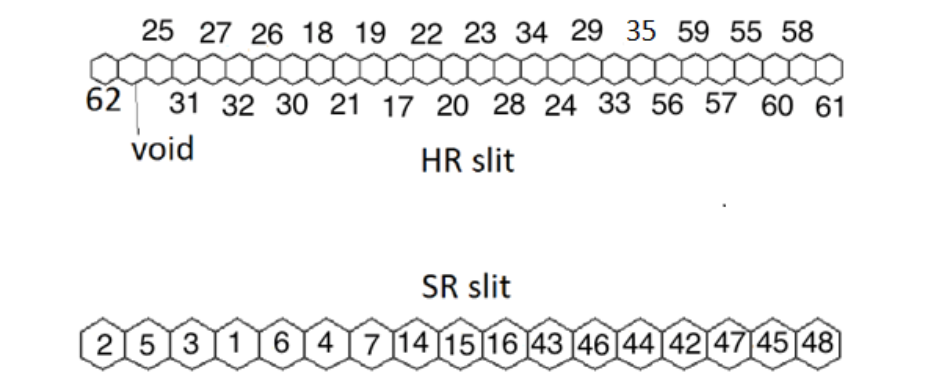}
   \end{center}
   \caption{ \label{fibernumbers} Mapping of the fibers from the IFUs to the slits. Fiber 62 is the internal ThXe calibration fiber. }
\end{figure}

The mapping of the fibers between the input of the IFUs and the corresponding pseudo-slits is carefully chosen, and is illustrated in Figure~\ref{fibernumbers}. For the standard resolution slit, the sky fibers are located between the two sets of object fibers, which prevents any cross contamination of the object spectra with each other. For the high resolution mode, the sky fibers are at one end of the slit. The internal calibration fiber (fiber 62 in Figure~\ref{fibernumbers}) is at the other end of the slit, close to the object fibers.

For each set of fibers corresponding to a specific IFU, there are no spaces between the fibers in the detector plane. At fixed spectrograph size (i.e. cost), leaving no space between fibers substantially improves etendue and hence sensitivity, which is a primary goal of GHOST. It enables highly binned and highly sensitive modes that are not as affected by detector dark current as they would be if there were gaps between fibers. Critically, it must be remembered that GHOST is a multi-fiber spectrograph where all fibers look at the same object, and therefore cross-talk between fibers does not result in a degradation in science performance as it would for a multi-fiber spectrograph where each fiber looks at a different object. For the data reduction (Section 5), there are no existing or currently planned data reduction modes that extract individual fibers.

More important than the relative locations of the object, sky and calibration fibers on the slit, however, is the internal mapping of the object fibers. These are chosen such that the fiber that is at the center of the IFU is at the center of the slit, those fibers that are on the outer ring of the IFU are on the outside of the slit and, in the case of the high resolution bundle, those that are in the middle of the ring are between the inner and outer fibers on the slit. The result is that during good seeing the flux in the object spectra become very concentrated in just a few rows on the detector, while in poor seeing the light is more evenly distributed across the slit. In Figure~\ref{fibernumbers}, fiber 62 is the simultaneous ThXe fiber.

\subsubsection{Throughput of optics}

\begin{figure}
   \begin{center}
   \includegraphics[width=8cm]{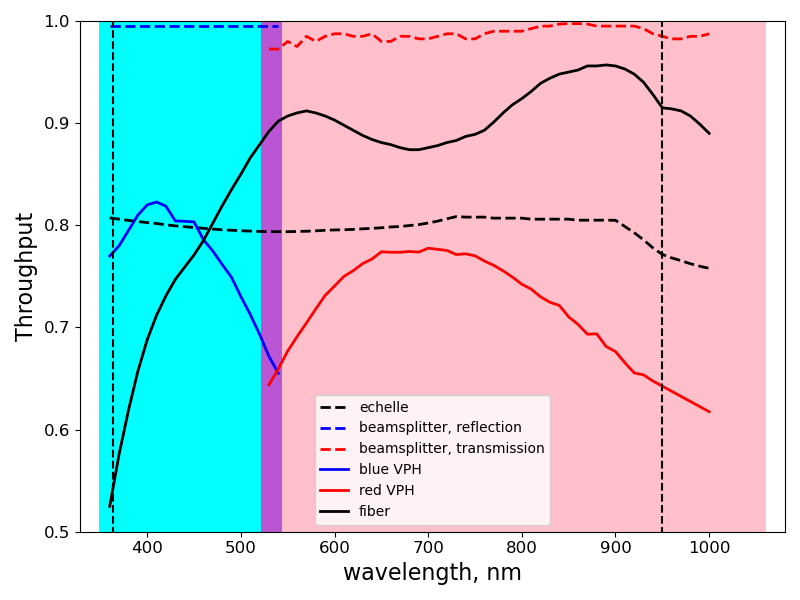}
   \end{center}
   \caption{ \label{componentthroughput} Measured throughputs of major optical components for GHOST. The cyan and pink regions show all wavelengths that fall on the blue and red detectors, respectively. The purple region is the transition region of the beam-splitter, where these wavelengths appear on both detectors. The dashed vertical lines correspond to the wavelength region of GHOST over which the design was optimised.}
\end{figure} 

GHOST achieves high throughput through the careful selection and optimization of dispersers, optics and coatings. Advances in fibers and coating technology have allowed this fiber-fed instrument to have a throughput competitive with Cassegrain-mounted instruments but with the advantage of the highly stable pier lab environment for the instrument, a worthwhile trade-off in design. 

Following the light through the GHOST system:

\begin{itemize}
    \item the {\it Atmospheric Dispersion Corrector} is an innovative mini-ADC design which allows thinner prisms and higher blue throughput (see Section~\ref{sec:adc});
    \item the {\it 31.6m fiber} is the Polymicro FBPI, a fiber that via specialized processing has the red attenuation of a low OH fiber and the blue attenuation of a high OH fiber;
    \item the {\it {\'e}chelle} (Richardson Grating Labs MR234) has near theoretical diffraction efficiencies of $\sim 80\%$ at order centers, a performance as good or better than the legendary MR160 R4 grating;
    \item the {\it white pupil relay optics}, which have four reflections in the optical path, are coated with Quantum Coatings UV350AG silver coating. This coating is a protected silver coating with UV enhancement whose performance exceeds that of aluminum at 360nm and is equivalent to bare silver above 480nm;
    \item the {\it beamsplitter}, manufactured by Asahi Spectra, has an extraordinary  $99.5\%$ reflectivity into the blue arm, and $>97.5\%$ (with an average of $99\%$) transmission into the red arm;
    \item the {\it VPHs}, manufactured by Kaiser Optical Systems, are slant groove gratings, allowing high efficiency in a non-Littrow configuration (necessary to control grating ghosts in the system). The final gratings efficiencies are near ideal for the system as the blue VPH efficiency was ``tilted'' to the blue end, exactly where extra performance was desired, and the red VPH has a broader efficiency curve compared to the theoretical curve, and again considered ideal for GHOST science performance. 
    \item {\it Glass and coating optimization} has minimized losses in the cameras, with each camera's transmittance designed to be greater than 96\% and 94\% for the red and blue cameras respectively. 
\end{itemize}

Performance curves for key components are shown in Figure~\ref{componentthroughput}. Here, the cyan and pink regions show all wavelengths that fall on the blue and red detectors, respectively. The purple region is the transition region of the beam-splitter, where these wavelengths appear on both detectors. The dashed vertical lines correspond to the wavelength region of GHOST over which the design was optimised. The overall throughput of the entire system and the overall sensitivity of GHOST is discussed in detail in Section~\ref{sec:sensitivity}.

\subsubsection{Spectrograph image quality}

\begin{figure*}
   \begin{center}
   \includegraphics[width=16cm]{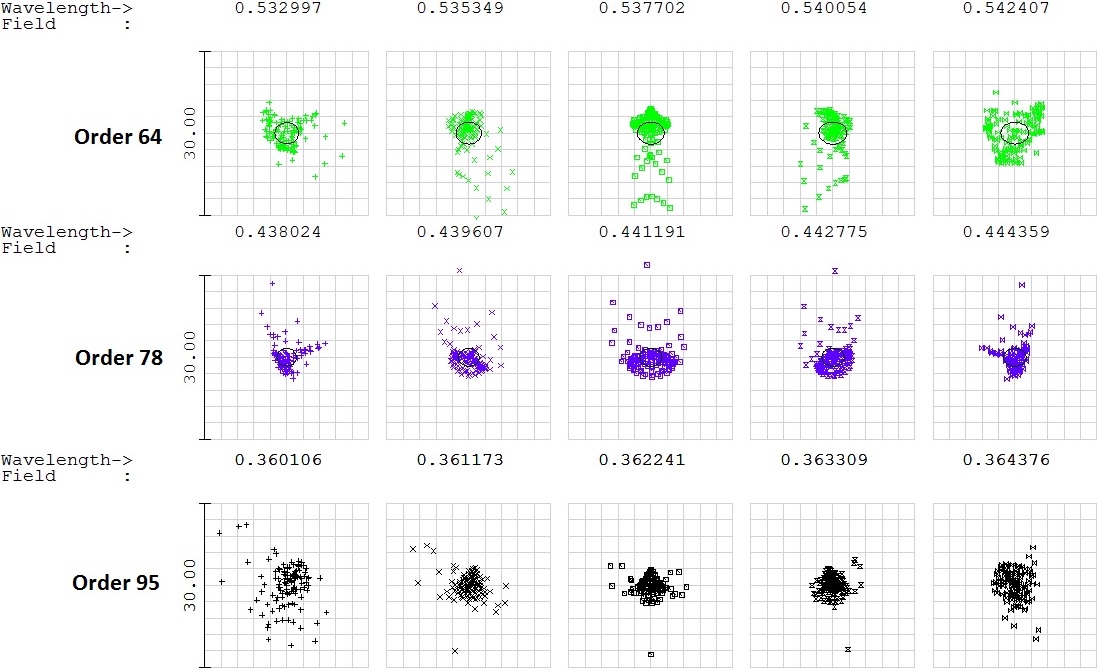}
   \includegraphics[width=16cm]{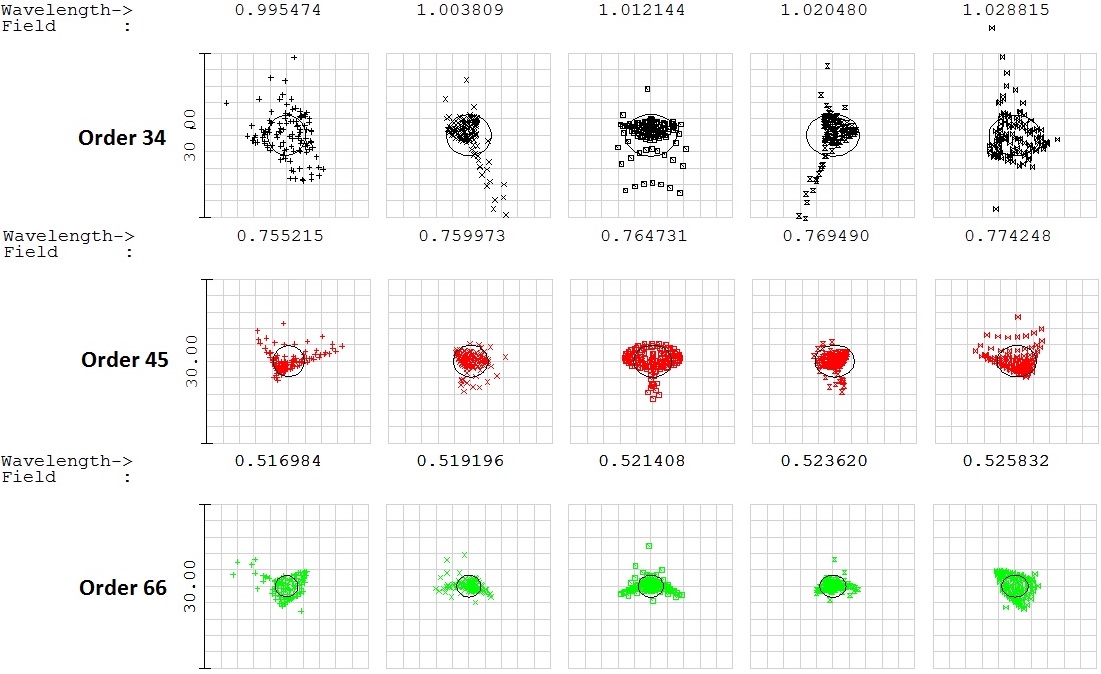}
   \end{center}
   \caption{\label{spotdiagram} Spot diagrams for the blue camera (left panel) and red camera (right panel). Overall box sizes are 2 pixels (30$\mu$m) on a side. Wavelengths in microns are given below each spot diagram.}
\end{figure*} 

GHOST was designed in order to achieve excellent image quality from $363 - 950$\,nm. For example, the zero-Petzval sum relay introduces no grating aberrations, eliminating the need for a mitigating effort and greatly improving the image quality and depth of focus of the white pupil relay/cross-disperser system over the alternative solutions of cylindrical power on the fold mirror or a cylindrical field lens. The blue and red camera designs have also been optimized with image quality in mind. Theoretical spot diagrams for the blue camera (left panel) and red  camera (right panel) are shown in Figure~\ref{spotdiagram} for the top, middle and bottom orders in each camera. The overall box size in the spot diagram is $2 \times 2$ pixels, with grid cells equivalent to 0.2 pixels. At each wavelength the circles represent the diffraction limit, and in some cases these are elliptical due to beam anamorphism. It can be seen that the design is close to the diffraction limit for the red orders. 

The actual delivered image quality was examined after integration of the entire bench spectrograph by taking multiple exposures of a ThXe source injected into the system with a 10$\mu$m fiber. Given that the pixels of the detector are 15$\mu$m, this is equivalent to a point source for GHOST. For each arm of the spectrograph, images were bias-subtracted and median-combined.

The majority of the arc lines do not, of course, land perfectly centered on a pixel. But given the very large number of lines, there will be some that do land on the detectors near a pixel center. An algorithm was constructed that looked for peaks in the counts and compared these to the surrounding pixels. Via this algorithm, many hundreds of cut-outs (11 x 11 pixels) were generated for lines which appeared to have good image quality (i.e., where the counts in the central pixel were a high fraction of the total counts in the neighbouring pixels). These cut-outs were then studied more carefully. Specifically, they were background subtracted (using the median of the counts in nearby pixels but away from the track of the spectrum) and continuum subtracted (using the median of the counts in each row in the vicinity of each line, but avoiding the line itself). The ensquared energy for each line is then estimated as the ratio of the flux in the central pixel compared to the sum of the flux in the inner $5 \times 5$ region of the cutout.

After checking for spurious values (e.g., where there was another bright line in the cut-out region), we rank-order the lines in order of decreasing ensquared energy. With many lines identified, we expect the highest ensquared energy values to correspond to those lines that are, coincidentally, well centered in the pixel. 

\begin{figure*}
   \begin{center}
   \includegraphics[width=12cm]{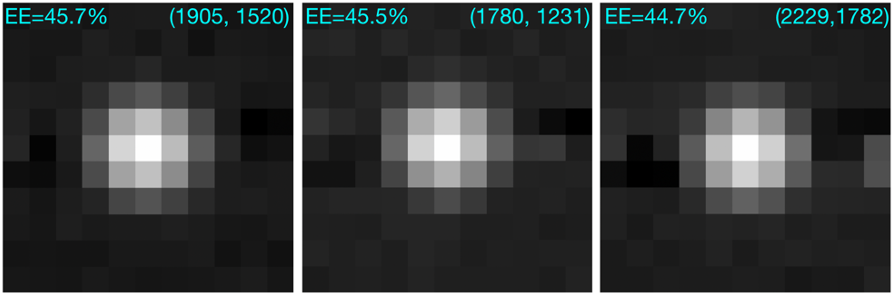}
   \includegraphics[width=12cm]{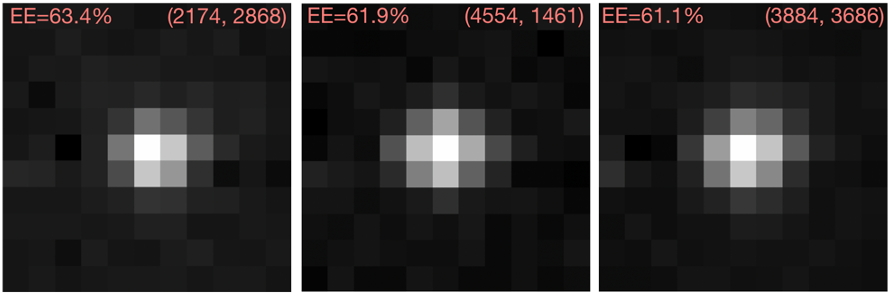}
   \end{center}
   \caption{\label{10micron} Top row: the three ThXe lines from the blue arm with the highest ensquared energy in a single pixel, using a 10$\mu$m fiber feed. Bottom row: same, but for the red arm.}
\end{figure*} 

The top row of Figure~\ref{10micron} shows the 3 lines with highest ensquared energy for the blue detector, and the bottom row of the same figure shows the 3 lines with highest ensquared energy for the red detector. Coordinates correspond to the x-y pixel locations of these lines in the original analysis.  We conclude that the ensquared energy for the blue arm of GHOST is $44 – 46$\% in a single pixel. The ensquared energy for the red arm of GHOST is $61 – 64$\% in a single pixel. Alternatively, for the blue arm, the delivered image quality translates into an ensquared energy of greater than 95\% of the flux within the central 3 pixels; for the red arm, the delivered image quality translates into an ensquared energy of approximately 98\% of the flux within the central 3 pixels. 

\subsubsection{Large format detectors}

\label{sec:detectors}

\begin{figure}
   \begin{center}
   \includegraphics[width=\columnwidth]{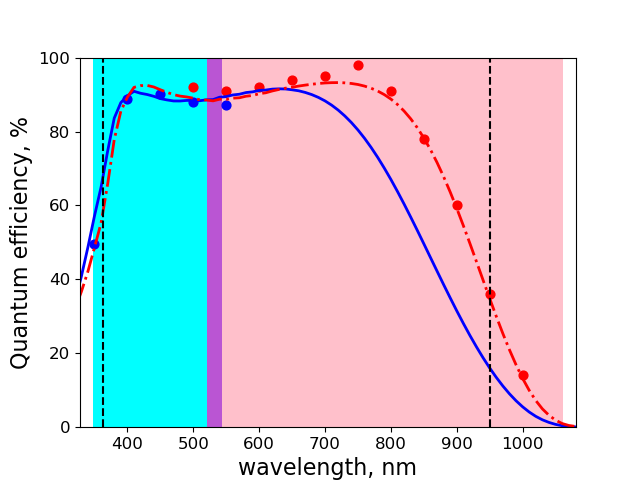}
   \end{center}
   \caption{\label{qe} Quantum efficiency of the blue and red detectors (blue solid and red dot-dashed lines respectively). The manufacturer performance estimates are shown as lines, and the measured point for our specific devices as shown as points.}
\end{figure} 

\begin{table*}
\begin{center}
\begin{tabular*}{0.75\textwidth}{l|ccc|ccc}
& \multicolumn{3}{c|}{Blue} & \multicolumn{3}{c}{Red}\\
\hline\\

Model & \multicolumn{3}{c|}{e2v CCD231-84, standard silicon} & \multicolumn{3}{c}{e2v CCD231-C6, deep depletion}\\
AR Coating &\multicolumn{3}{c|}{thin astro multi-2}& \multicolumn{3}{c}{astro multi-2}\\
Number of pixels & \multicolumn{3}{c|}{$4096 \times 4112$}&\multicolumn{3}{c}{$6114 \times 6160$}\\
Pixel size & \multicolumn{3}{c|}{$15\,\mu$m} & \multicolumn{3}{c}{$15\,\mu$m} \\
Orders & \multicolumn{3}{c|}{$m = (98)~95 - 64$}&\multicolumn{3}{c}{ $m = 66 - 34~(32)$}\\
Wavelength range& \multicolumn{3}{c|}{$(348)~363 - 544$\,nm}&\multicolumn{3}{c}{$ 521 - 950~(1061)$\,nm}\\
Read-out modes& {\it Slow} & Medium & Fast & Slow & {\it Medium} & Fast\\
Read-out time& 50s & 25s & 10s & 100s & 55s & 22s\\
Read-out noise& $2.1e-$ & $2.5e-$ & $4.7e-$ & $2.2e-$ & $2.4e-$ & $4.5e-$\\
Nominal Gain e-/ADU & \multicolumn{2}{c}{0.55} & 0.73 &\multicolumn{2}{c}{0.5} & 0.7\\
Dark current& \multicolumn{3}{c|}{$0.9e-$/pixel/hour}& \multicolumn{3}{c}{$0.4e-$/pixel/hour}\\
  Operating temperatures & \multicolumn{3}{c|}{$-115^\circ$C }& \multicolumn{3}{c}{$-110^\circ$C }\\
  Detector output &  \multicolumn{3}{c|}{ 4 channel x 16 bits} &  \multicolumn{3}{c}{ 4 channel x 16 bits} \\
\hline\\
\end{tabular*}
\caption{Summary of the main characteristics of the GHOST detectors. Readout modes in italics are the default modes for the detectors. Orders and wavelengths in parentheses correspond to the orders and wavelengths that fall on the detector, whereas the values outside of parentheses correspond to the values for which the design is optimised.}\label{tab:detectors}
\end{center} 
\end{table*}

GHOST uses a 2-channel optical design to deliver high spectral resolutions with a vast wavelength range while remaining at or close to critically sampled. This has been made possible by large-format e2v detectors that were not available during the early years of the GHOST design, but which were expected to become available prior to fabrication. 

The detectors delivered by e2v exceeded expectations. The blue channel uses an e2v CCD231-84 detector, which has 4096(H) x 4112(V) 15$\mu$m pixels, is standard thickness silicon, and which has  a thin-astro-multi2 anti-reflection coating. The red channel uses an e2v CCD231-C6 detector, which has 6144(H) x 6160(V) 15$\mu$m pixels, uses deep-depletion silicon with the anti-fringing option, and which has an astro-multi2 anti-reflection coating. The main characteristics of the detectors relevant from a science user perspective are summarised in Table~\ref{tab:detectors}. We note that, for high resolution spectroscopy, fringing is not usually considered critical. For example, for 40 micron thickness and a refractive index of 3.6, the expected fringe spacing is $28\AA$ at a wavelength of $9000\AA$. This equates to $\sim 250$ spectral pixels in GHOST. For most high resolution science, these 1\% amplitude fringes should flat field out, and can additionally be removed by tweaking the continuum normalisation.

The coatings on both detectors are a significant improvement over the single layer anti-reflection coatings previously available over the wavelength ranges of relevance for GHOST. Figure~\ref{qe} shows the manufacturer estimate of performance  including the thin-astro-multi2 coating on standard silicon (blue detector, blue solid line) and the astro-multi2 coating on deep depletion (red detector, red dot-dashed line). Points show the actual delivered quantum efficiency (QE) for our specific devices from e2v test results. For the QE, as well as the other characteristics discussed below, the devices delivered by e2v meet or exceed our performance expectations.

Each detector is housed in a custom-built vacuum cryostat, operated by a 4-channel ARC GenIII controller, and cooled with an ARS closed-cycle cryocooler to below $-100^\circ$C.  Thermal control of each detector is maintained by a Lakeshore temperature controller, and the current operating temperatures of the blue and red detectors are $-115^\circ$C and $-110^\circ$C, respectively.

In addition to large format and high QE over a broad spectral range, an additional requirement for the detectors was to have very low read noise ($< 3e-$) and low dark current. Both detectors can be read out at one of three different speeds (slow, medium and fast), with read-noise increasing as speed increases. The default modes are slow readout for blue and medium readout for red, in recognition that read noise is usually more important at photon-starved blue wavelengths, and that this setting allows the much larger red detector to be read out in nearly the same time as the blue detector. In these default modes, both detectors take around 50\,seconds to read out. The read-noise in the blue is measured to be $2.0 - 2.1 e$- with a gain of $0.52 - 0.59 e$-$/DN$ for the four output channels. In the red it is measured to be $2.3 - 2.5 e$-, with a gain of $0.5 - 0.55 e$-$/DN$ for the four outputs. 

Dark current has been measured and is impressively low. Indeed, open shutter darks (which take into account both the intrinsic dark current of the detector as well as stray light from the background) show counts in the blue corresponding to 0.9e-/pixel/hour and in the red of 0.4e-/pixel/hour. As well as demonstrating the high quality of the detector, these numbers are also the result of considerable care that was taken during integration at Gemini South to minimise or block any and all identifiable sources of light leaks in the system.

\subsubsection{Slit tilt}
\begin{figure*}
   \begin{center}
   \includegraphics[width=\textwidth]{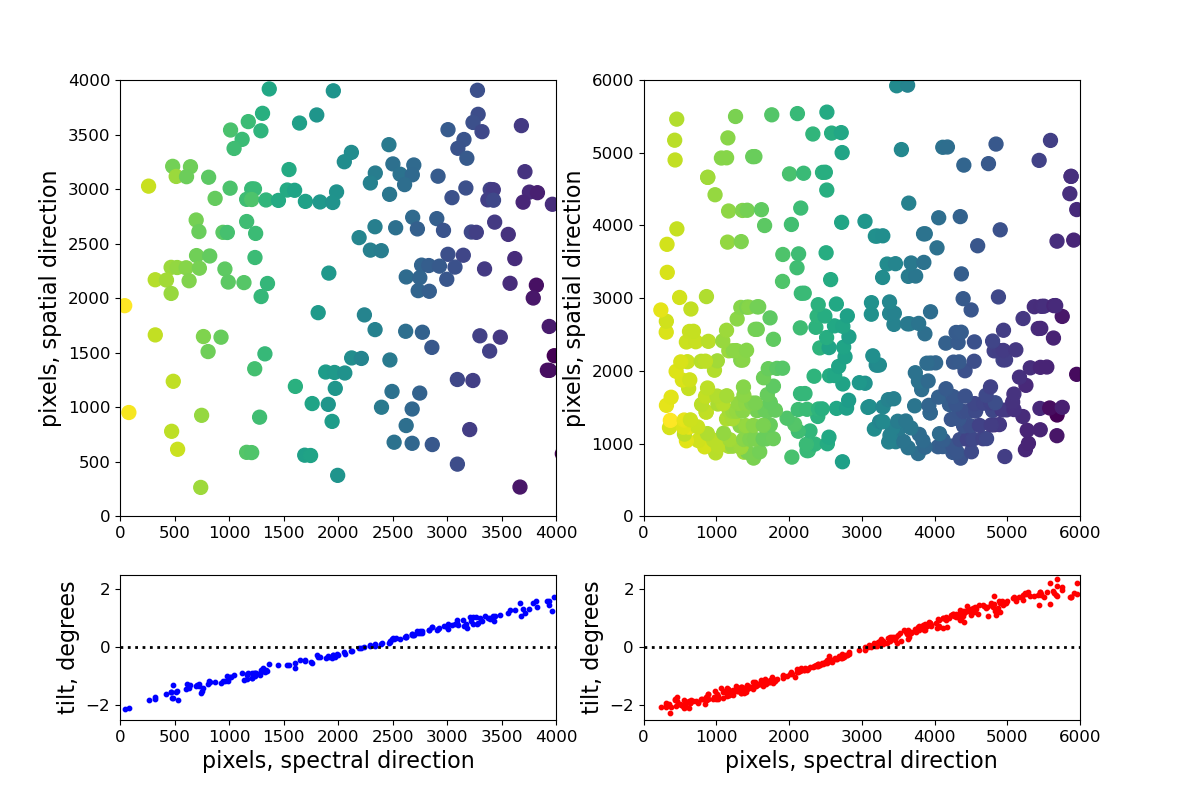}
   \end{center}
   \caption{\label{rotate_slit} Slit tilt as a function of 2D position on the detector (top panels) and as a function of x-pixel position (bottom panels), for the blue (left) and red (right) detectors. The color scheme in the top panels shows the slit tilt, from -2 degrees (yellow) to +2 degrees (purple).}
\end{figure*} 

GHOST is an {\'e}chelle spectrograph. A generic feature of such spectrographs is that the slit is not vertical with respect to the columns (spatial direction) of the detectors, but instead rotates with wavelength. The changing alignment of the slit with the detector columns as a function of wavelength necessitates that the spectra are resampled and the data interpolated in order to successfully extract the spectra.

The slit of GHOST also changes alignment with respect to the detector columns as a function of wavelength, but the net rotation has been minimised. This has been achieved by optimising the tilts from individual components such that they have a net cancelation effect. Figure~\ref{rotate_slit} shows the tilt as a function of 2D position on the detector (top panels) and as a function of x-pixel position (bottom panels), for the blue (left) and red (right) detectors. These tilts are empirical measurements from a set of arc line observed in standard resolution, where the full slit was illuminated. The physical tilt of the slits is the same in both high and standard resolution modes, and they are essentially vertical close to the center of the detector, and have a maximum of a 2 degree tilt either side of vertical at the chip edges. This small tilt is accounted for in the data reduction pipeline.

\subsubsection{Thermal enclosure}

\begin{figure*}
   \begin{center}
   \includegraphics[width=10cm]{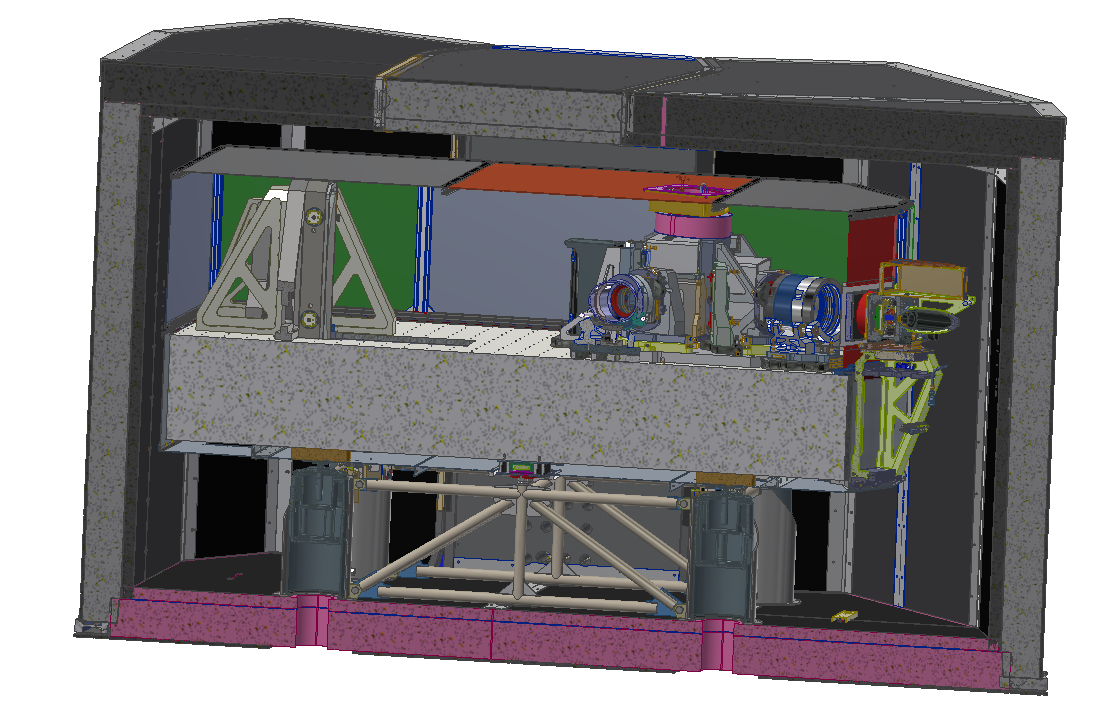}
   \includegraphics[width=10cm]{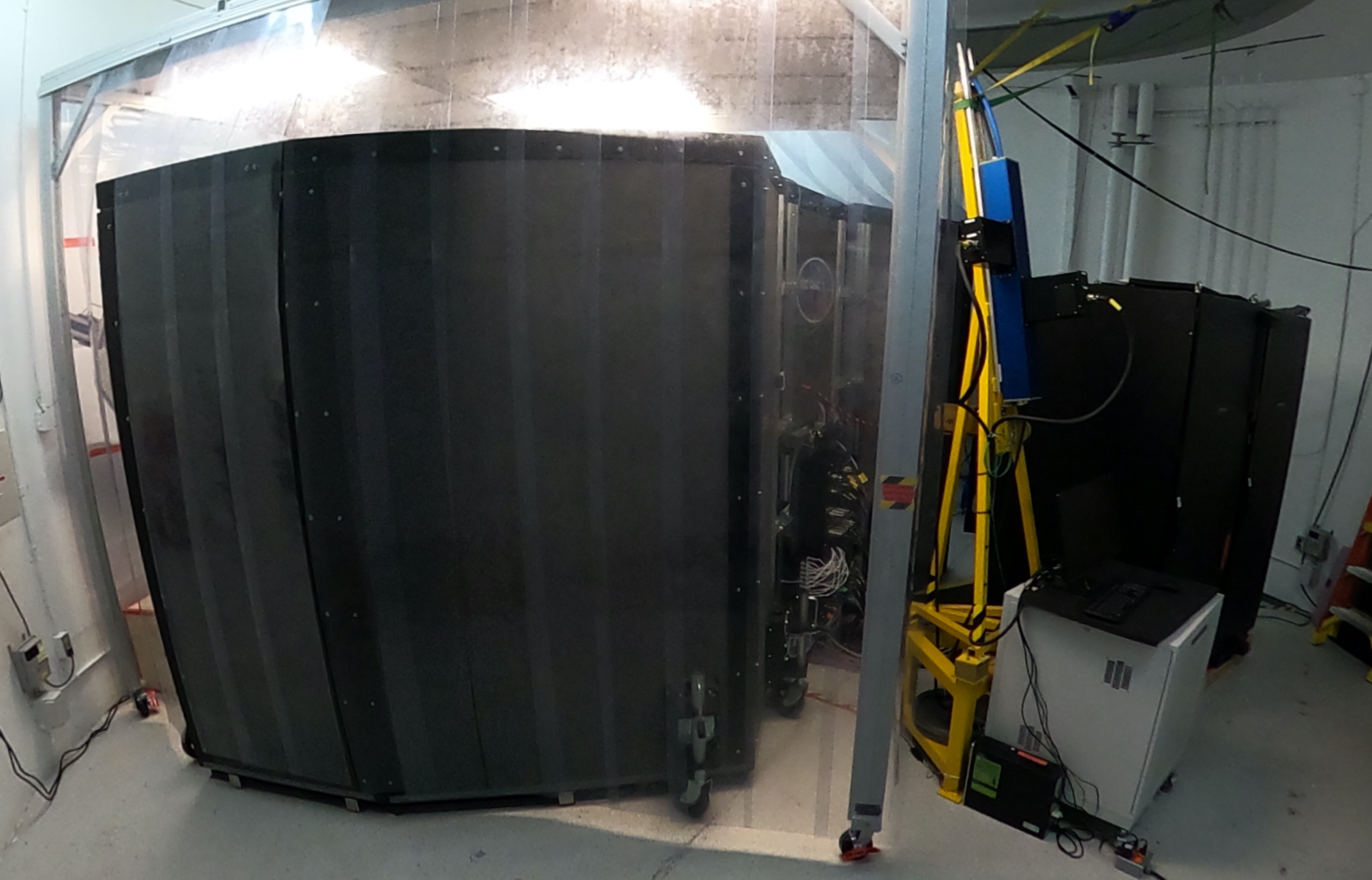}
   \end{center}
   \caption{\label{enclosure} Top panel: cut away of the outer enclosure showing the GHOST optical bench, the inner shell and the outer enclosure. Bottom panel: wide-angle shot of GHOST, in its enclosure, in the pier lab of Gemini South.}
\end{figure*} 

GHOST is located in the pier-lab of Gemini South, which is a quiet, dark and passive environment. However, great additional care has been taken to ensure that GHOST's environment is temperature stable and vibrationally stable. The left panel of Figure~\ref{enclosure} shows that the optical bench is surrounded by an inner enclosure (shell) that helps protect the optics and block stray light. The inner enclosure is then surrounded by an outer enclosure. This outer enclosure consists of 20 heated thermal panels forming an encompassing structure with a stationary ‘bridge’ assembly and two removable sections for access. The outer enclosure provides a temperature-stable, dark environment for the bench spectrograph, and the legs of the bench provide vibrational isolation for the system.

\begin{figure*}
   \begin{center}
   \includegraphics[width=\textwidth]{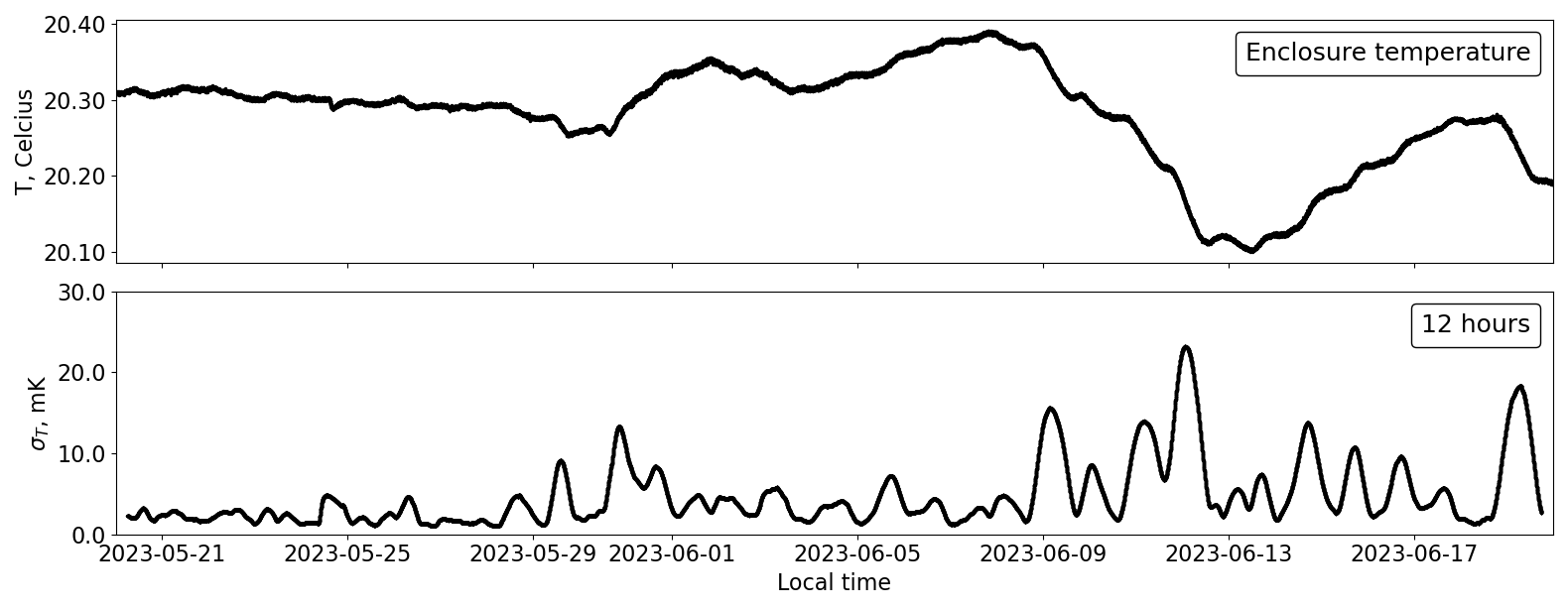}
   \end{center}
   \caption{\label{TempStability} The top panel shows the readings from one of the thermistors monitoring the inner enclosure over a recent period of 1 month. The second panel shows the standard deviation in the inner enclosure temperature estimated in running windows of 12 hours.}
\end{figure*} 

The thermal performance of the enclosure is  monitored using an array of sensors, including NTC thermistors embedded in the heating plate of each panel ($\pm 0.005^\circ$C), and NTC temperature sensors that monitor air, bench and grating temperatures while the instrument is in operation ($\pm 0.01^\circ$C).

At the time of writing, the temperature control of the enclosure is still being fine-tuned, although the short-term temperature stability of the enclosure is still good. The top panel of Figure~\ref{TempStability} shows the readings from one of the NTC temperature sensors (monitoring the inner enclosure) over a recent period of 1 month (June 2023; note that the thermistors all track each other very closely). The second panel aims to quantify the stability on 12 hour timescales by showing the standard deviation in a running window of 12 hours duration.  Clearly, the temperature is stable on scales of order 10 - 20 mK.

Once the temperature control of the enclosure is operating nominally, it is the design expectation that the current level of performance observed on short timescales will be possible on longer duration timescales as well (at least a few nights), and that the short-timescale performance should also be improved to $\lesssim 10$\,mK. It is worth noting that this temperature performance is being achieved without the use of pressure vessels or cryogenic cooling. Rather, this novel thermal enclosure is designed in a way to allow personnel access to the bench spectrograph while maintaining a suitable footprint relative to spectrograph size. The outer enclosure is able to warm from ambient conditions to a stabilized environment in under 24 hours. We expect that this highly stable temperature environment for GHOST will be of great utility in performing precision spectroscopic measurements.

\section{Key operational characteristics}

\subsection{Focal plane considerations}

\label{sec:fp}

\subsubsection{On-sky layout and IFU patrol fields}

\begin{figure*}
   \begin{center}
   \includegraphics[width=12cm]{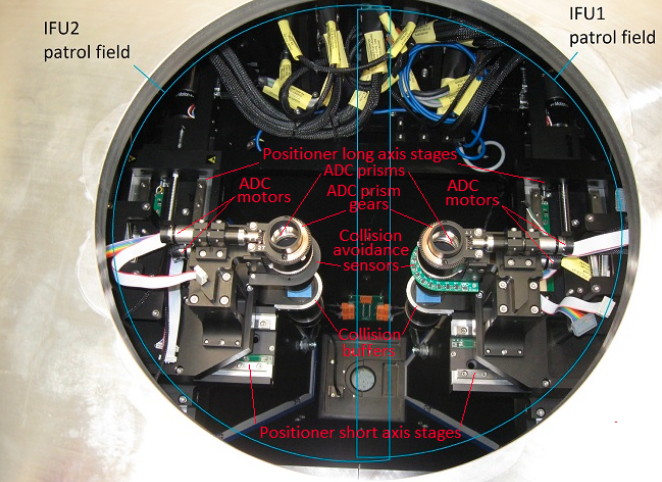}
   \end{center}
   \caption{ \label{patrolfield} GHOST Cassegrain unit viewed through the aperture during assembly and before installation of the telecentricity lens.}
\end{figure*}

The distinction between standard and high resolution mode in GHOST is a distinction between the size of the microlenses that define the entrance aperture of the fibers at the focal plane. Both modes sample a 1.2" diameter aperture on the image plane, the size of which is chosen to sample $70\%$-ile seeing conditions at Gemini South ($\sim 0.8$"). However, in standard resolution mode, the microlenses effectively slice this by a factor of 3, and in high resolution, they slice this by a factor of 5. 

Figure~\ref{microlens_in} shows the on-sky projection of the IFUs. Both IFU heads contain standard and high resolution microlenses arranged to sample targets and/or sky. IFU1 contains  a standard resolution target bundle, the standard resolution sky bundle, and the high resolution target bundle. IFU2 contains a second standard resolution target bundle, and the high resolution sky bundle. Within a given IFU, the locations and orientations of the standard resolution and high resolution microlens arrangements are fixed relative to each other.  In practise, the configuration shown in Figure~\ref{microlens_in} means that a single scientific target can be observed per exposure at high resolution using IFU1. In standard resolution, up to two science objects can be observed simultaneously. 

Figure~\ref{patrolfield} shows a photograph of the IFUs in the Cassegrain unit. Each IFU head is mounted on an X -- Y stage, and together they patrol a 7.5 arcmin field of regard. Specifically, each IFU head patrols one semi-circle of the circular field as outlined in blue in Figure~\ref{patrolfield}, with each IFU head able to reach into the other semi-circle by 5.5mm  (9 arcsecs). During commissioning, a pair of targets were acquired that were 7.45 arcminutes apart, and this should be considered the absolute maximum separation of targets in any single observation allowable with the GHOST hardware.

The IFU heads are equipped with collision avoidance Hall sensors to prevent them colliding with each other. During commissioning, experimentation showed that this limits the closest approach of the IFU heads to 102 arcseconds when approaching each other along the X or Y axis, and 123 arcseconds when approaching each other on a diagonal. Targets closer than these separations cannot be observed simultaneously in a single exposure.

Any time two objects are being observed simultaneously, it is highly recommended that they are placed symmetrically relative to the field of regard. The IFU heads move in a plane, but the Gemini focal surface is curved. Therefore, there is a change in focus with radius. A focus offset is applied for any observation where the targets are not at the center of the field, corresponding to the necessary offset given the distance of the objects from the center of the field. If the objects are not symmetrically placed, however, then at least one of the objects will be slightly out-of focus, leading to small but unnecessary additional losses at the IFU injection.

\subsubsection{Sky subtraction strategies}

In high resolution mode, the only option for obtaining a dedicated sky observation is to use the high resolution sky fibers. Since these are located on IFU2, then the sky position can be anywhere in the field of regard of IFU2. In standard resolution, a ``sky'' observation will be obtained anytime IFU1 is in use. These sky fibers are in a fixed location relative to IFU1 in the image plane (separated by 2.35 arcsecs) to the right of the science fibers (see Figure~\ref{microlens_in}), and it is envisioned that a correct selection of the Cassegrain rotation will ensure that these fibers sample an empty patch of sky for most science targets.

If two objects are being observed in standard resolution, then the dedicated sky fibers are the only option to obtain dedicated sky spectra. If, however, only one science object is being observed, then the second IFU head (likely IFU2) can be moved anywhere in its field of regard to an empty patch of sky to obtain a sky observation. Indeed, if it is so desired, the beam switching technique can be used, whereby one IFU is initially on the target, the second on sky, and then for the next sub-exposure, the second IFU is placed on the target, and the first IFU is moved to a sky field. This can be repeated until the desired total exposure time is obtained. This has the theoretical benefit that the target and sky are observed through the same IFUs (although not at the same time), reducing some of the systematics associated with obtaining sky observations with fiber systems. Should precise sky subtraction be critical for the science goal, it is worth noting that yet another handle on the sky will be obtained from comparing the signal in the outer ring of fibers in each target bundle to the inner fiber, for any period of good or excellent image quality (IQ). In this situation, the object flux will be concentrated in the central fiber (the central few rows of pixels of the spectrum), while the outer ring of fibers (the outer rows of pixels of the spectrum) will have a significant contribution from sky.

\subsubsection{Atmospheric dispersion compensation}

\label{sec:adc}

\begin{figure*}
   \begin{center}
   \includegraphics[width=14cm]{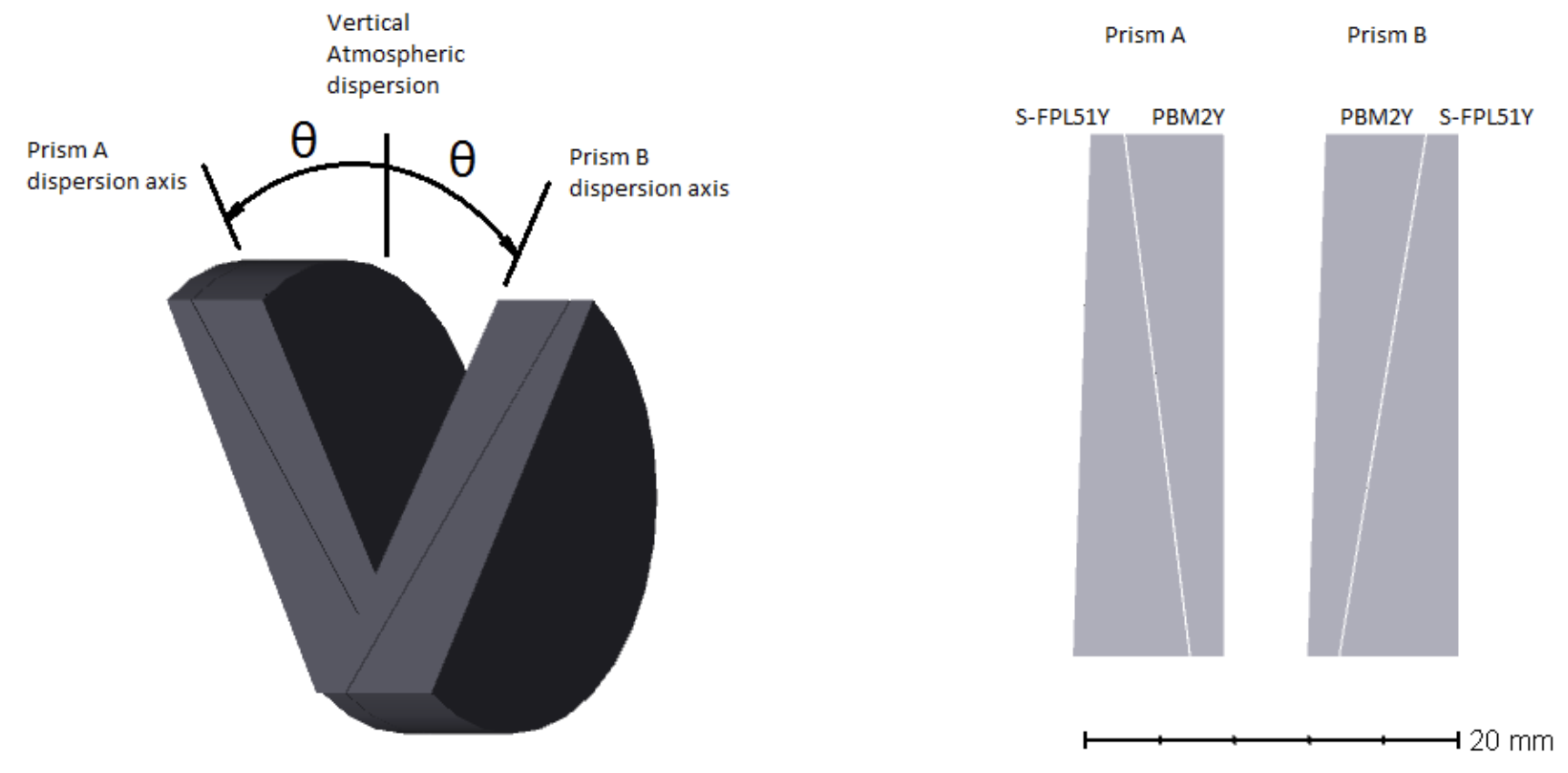}
   \end{center}
   \caption{ \label{adc_layout} Optical layout of GHOST atmospheric dispersion corrector.}
\end{figure*}

\begin{figure*}
   \begin{center}
   \includegraphics[width=8cm]{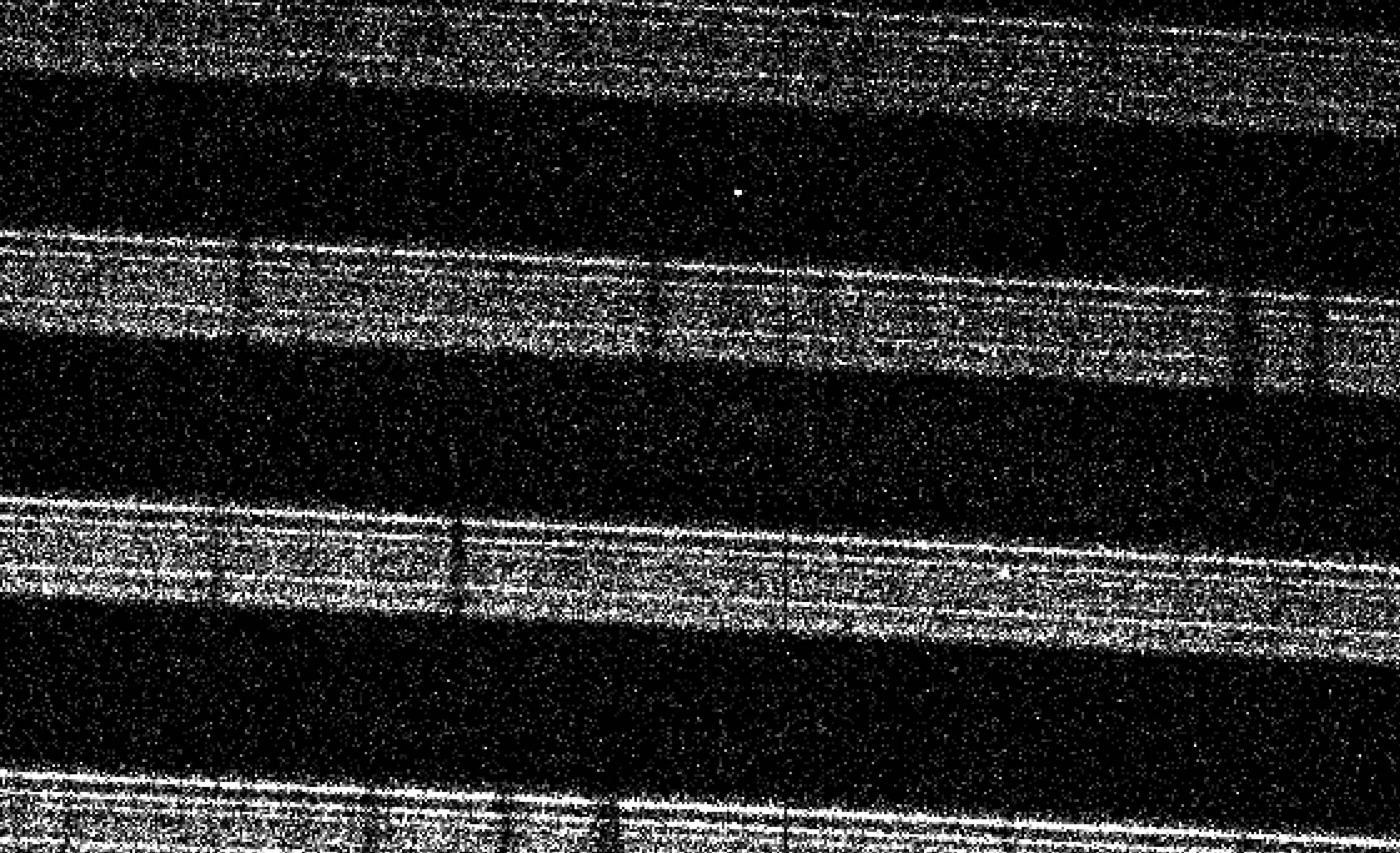}
   \includegraphics[width=8cm]{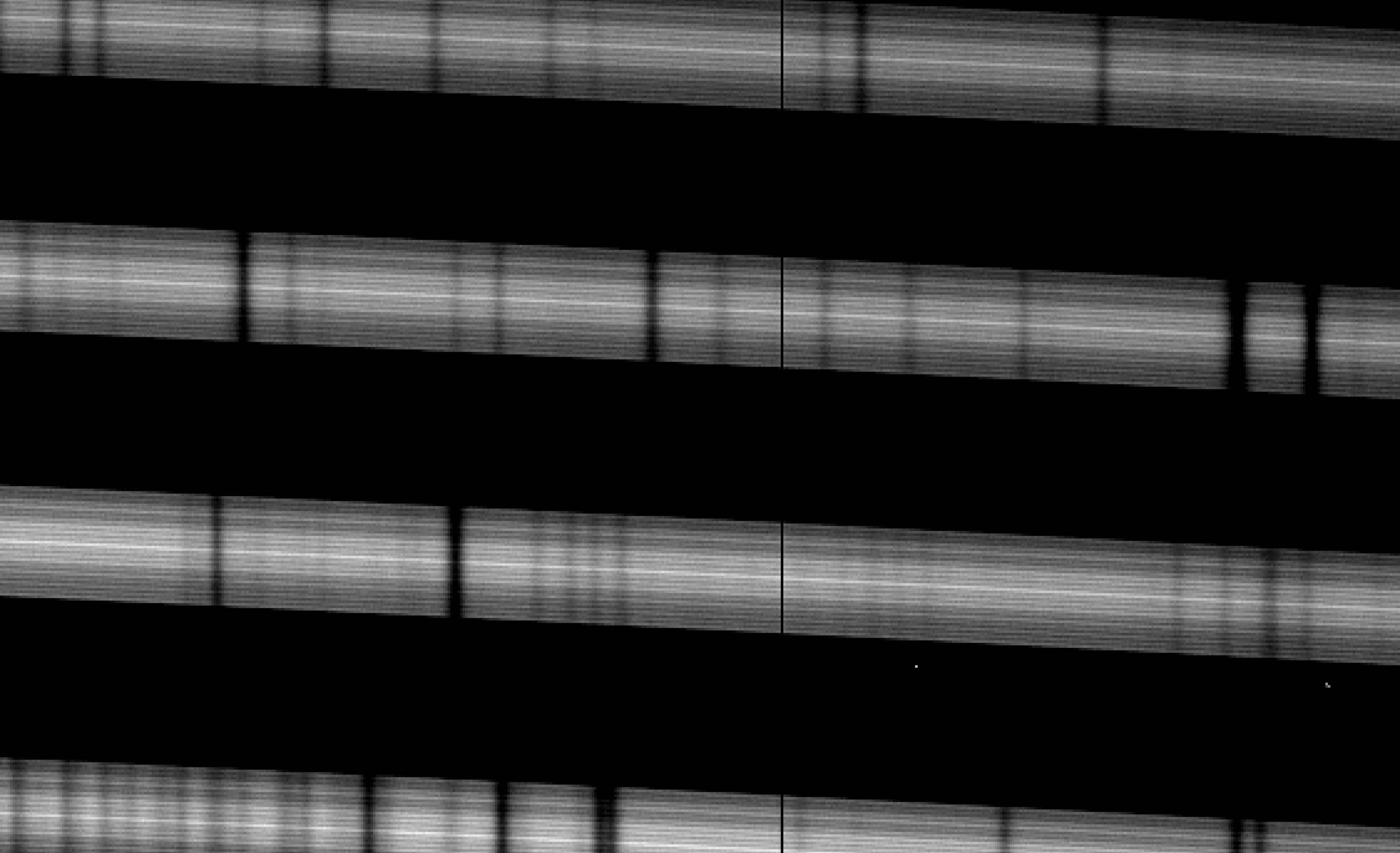}
   \end{center}
   \caption{ \label{adc} Left panel: a segment of an image from the blue detector showing a few {\'e}chelle orders for a source that was observed without the ADCs being switched on. Right panel: the same source observed with the same exposure time, but with the ADCs switched on.}
\end{figure*}

Critical to the scientific utility of most astronomical instruments is the ability to get light into the system. Given the entrance aperture of the IFUs is only 1.2", GHOST would be handicapped at this first hurdle if there was no correction for atmospheric dispersion: the dispersive effect of the atmosphere at an airmass of 1.5 means that an object's image at 350nm is displaced by 2.04 arcseconds relative to the image at 1050nm.

Each GHOST IFU has its own mini atmospheric dispersion corrector (ADC). Consequently, the prism thickness of the ADCs and their subsequent losses are much reduced compared to using a full-field ADC, while also allowing for a more precise correction. Each ADC is approximately 200mm upstream of the corresponding IFU, and is located after the telecentricity field lens. They are based on a Risley prism pair, and the design is such that they produce zero image shift: the bonded surfaces of the prisms deliver the necessary amount of dispersion while, at the same time, the angles of the air-glass surfaces are carefully chosen to maintain the path of the chief ray and minimize overall deviation at all ADC rotation angles. In the absence of such compensation, the boresight error would result in a zenith-distance-dependent image shift on the IFU and non-telecentric injection into the IFU optical cable. Figure~\ref{adc_layout} shows the configuration of the two counter-rotating cemented prism pairs of one ADC. At zero ADC rotation angle $\theta$, as shown in Figure~\ref{adc_layout}, the prisms’ dispersive power is oriented along the vertical in the opposite direction to the atmospheric dispersion, hence maximum compensation occurs. At non-zero angles, the horizontal components of the prism dispersion are mutually cancelled whereas the vertical ones are reduced proportionally to the magnitude of the atmospheric dispersion. At an ADC rotation angle of 90$^\circ$, the dispersing power of the prisms is nulled, for objects near the zenith.

The ADCs are absolutely essential to the successful operation of GHOST, and during commissioning they were tested as a function of airmass and were found to be operating nominally. They work to a zenith angle of 60 degrees, which is a hard limit set by the lens design; at even larger zenith angles, the ADCs will still apply the correction for 60 degrees, with a corresponding degradation in the suitability of the correction being applied. Figure~\ref{adc} shows the difference in the (2D) spectra obtained in the blue for when the ADCs are switched off (left panel) compared to when they are switched on (right panel). The content of this figure was fortuitously obtained as a result of an unplanned but illustrative oversight during commissioning.

\subsubsection{Aperture losses}

\label{sec:slitloss}

\begin{figure*}
   \begin{center}
   \includegraphics[width=8cm]{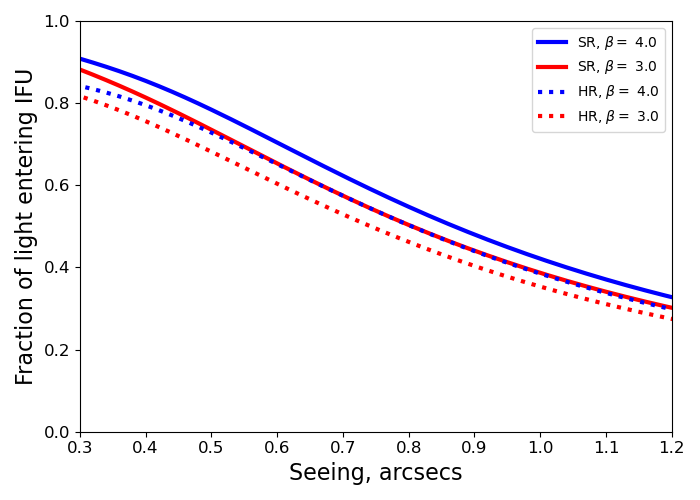}
   \includegraphics[width=8cm]{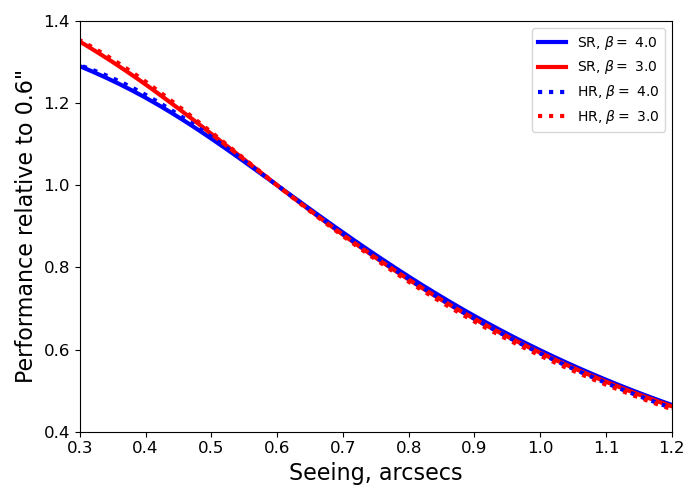}
   \end{center}
   \caption{ \label{slitloss} Left panel: fraction of light from a PSF that enters an IFU compared to the integrated flux as a function of seeing. Solid and dotted lines correspond to the aperture losses for the standard  and high resolution modes, respectively. Red lines correspond to a Moffat PSF with $\beta = 4$, and blue lines correspond to $\beta = 3$. Right panel: Same, but showing the fraction of light entering GHOST relative to performance at a seeing of 0.6".  These curves include the losses due to the fill-factor of the microlenses.}
\end{figure*}

Aperture losses refer to the light from a target that is lost to science because it does not make it into the instrument. For GHOST, this refers to the light in the point spread function (PSF) of a star that does not fall within the aperture defined by the microlens arrays. The geometry of these apertures is shown in Figure~\ref{microlens_in}. The microlenses in standard and high resolution also have a 93\%  and 85\% fill factor, respectively, due to the packaging.

For a PSF with a functional form with radius of $f(r)$ and a full-width-at-half-maximum $\theta$, we adopt a Moffat profile (\citealt{moffat1969}):

\begin{equation}
f(r) =  \left(\beta-1\right) / \left(\pi \alpha^2\right)\left[1+(r/\alpha)^2 \right]^{-\beta}
\end{equation}

\noindent where

\begin{equation}
\theta = 2 \alpha \sqrt{2^{1/\beta}-1}
\end{equation}

\cite{andersen2006} suggest that seeing limited PSFs are well fit by Moffat functions with a range of $\beta = 2.5 - 4.5$. \cite{racine1996} suggest $\beta = 4$, and $\beta = 3$ has been adopted for other systems, e.g., the Maunakea Spectroscopic Explorer, \cite{hill2018c} and references therein.  It is also likely that the exact value of $\beta$ to use will vary from telescope to telescope since it may depend on localized effects such as vibrations and dome seeing (\citealt{lai2019}).

Figure~\ref{slitloss} shows the aperture losses for the standard (solid lines) and high (dotted lines) resolution modes as a function of seeing, $\theta$, for a PSF that is perfectly centered in the IFU. Curves include losses due to the fill factor of the microlenses. Red lines correspond to a Moffat PSF with $\beta = 4$, and blue lines correspond to $\beta = 3$. The left panel shows the fraction of light that enters the IFU compared to the integrated flux of the PSF. Absolute aperture losses depend sensitively on the shape of the PSF: at $FWHM = 0.5”$, there are $> 6\%$ additional light losses for $\beta = 4$ compared to $\beta = 3$, and by 0.7”, there are 8\% additional light losses. 

At this early stage in the life of GHOST, it is unclear which (if either!) of these curves is the most appropriate for the instrument. However, over a continued period of GHOST operations (e.g., a year), it will likely be possible to build up a statistical understanding of the aperture losses as a function of seeing, given enough observations of spectrophotometric standards taken with good and stable sky transparency (where the differences in throughput after atmospheric extinction has been corrected for will correspond to differential aperture losses between observations). 

The steepness of the curves in the left panel of Figure~\ref{slitloss} is notable, and demonstrates that the total flux entering GHOST is quite sensitive to the seeing. The right panel of Figure~\ref{slitloss} makes this explicit, by showing the fraction of light entering GHOST relative to performance at a seeing of 0.6". Here, irrespective of the detailed functional form and resolution mode, it is apparent that a good rule of thumb is that $\sim 10\%$ more light enters GHOST for every 0.1 arcsec improvement in seeing. We also comment that the aperture losses are reasonably robust to centering errors: imperfectly centering the PSF by 0.1 arcsecs only causes a 1 – 2\% additional aperture losses, although this can grow to more like 6\% for PSF centering error of 0.2” and larger. 

Finally, the steep dependency between aperture losses and seeing provides an additional complication when aperture losses as a function of wavelength are considered. Since $\theta \propto \lambda^{-\frac{1}{5}}$ (e.g., see discussion in \citealt{woolf1982}), this means that the seeing near the blue end of GHOST (350nm)  is about 25\% larger than the seeing near the red end of GHOST (1050nm). For the case where the seeing in the r-band is 0.6", this corresponds to a seeing at [350, 1050nm] of [0.66", 0.54"] i.e, there is about 10\% more light entering GHOST at the reddest wavelengths than at the bluest wavelengths.

\subsection{On-detector considerations}

\subsubsection{Spectral format}

\begin{figure*}
   \begin{center}
   \includegraphics[width=8cm]{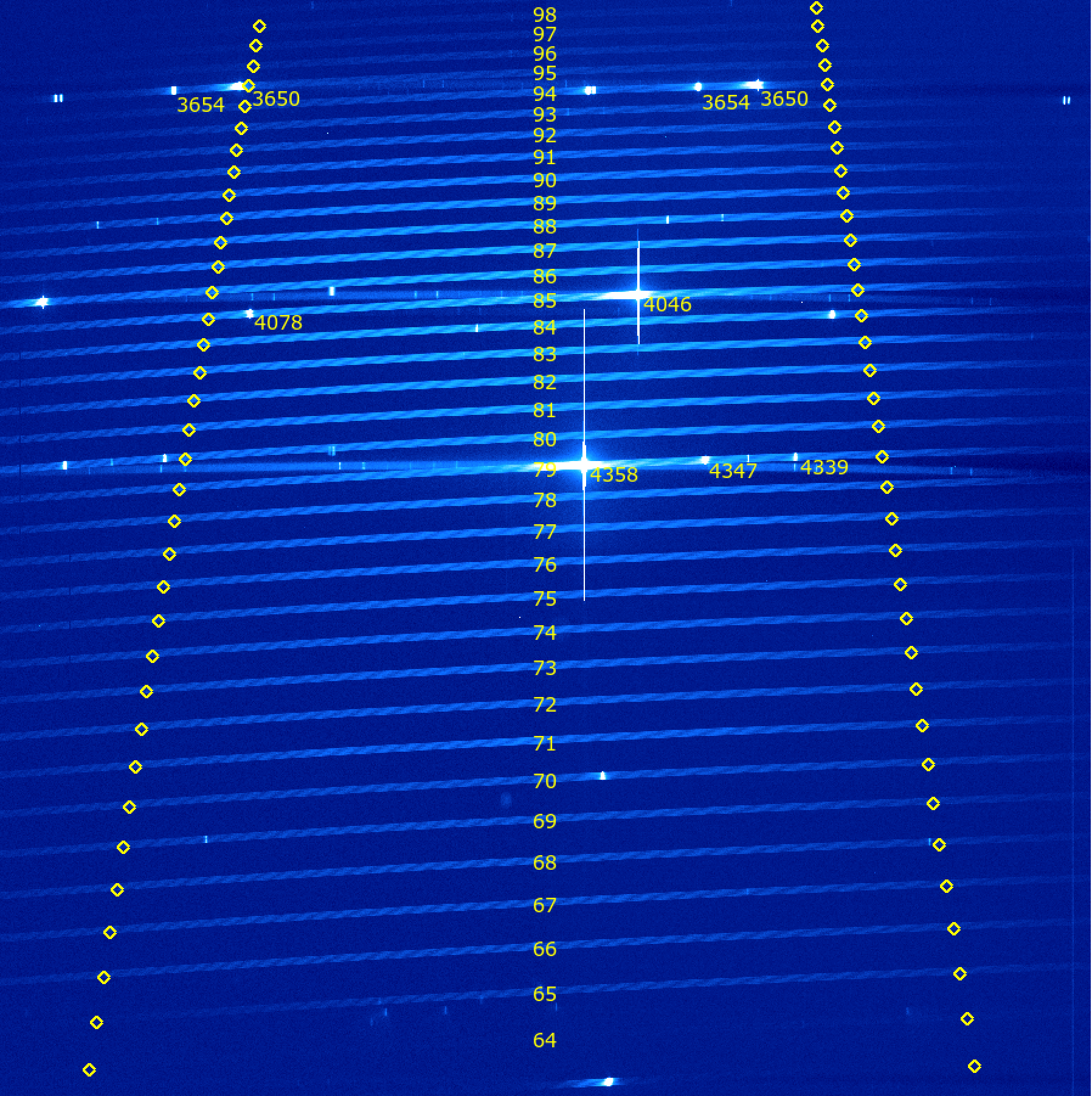}
   \includegraphics[width=8cm]{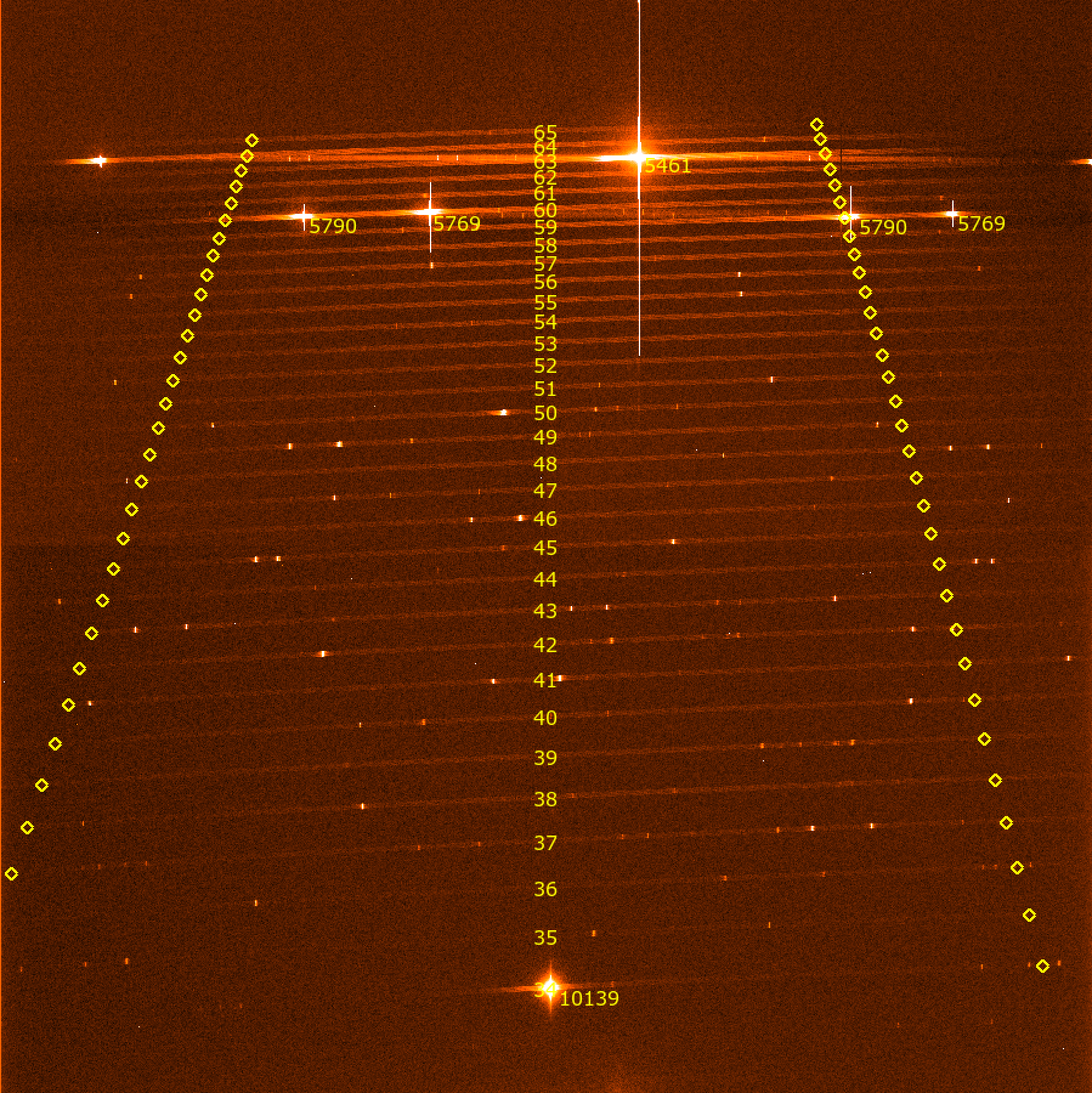}
   \end{center}
   \caption{ \label{hgspec} Full frame images from the blue (left panel) and red (right panel) detectors taken during integration at Gemini with light from a Hg lamp being fed down the standard resolution slit of IFU1. The wavelengths from some of the bright  Hg lines are marked. Each order is labelled with its number, and the circles near the ends of the orders indicate the free spectral range.}
\end{figure*}

Figure~\ref{hgspec} shows the global layout of the {\'e}chelle spectra on the blue (left) and red (right) detectors. These were obtained during final integration at Gemini by feeding light from a Hg lamp through the standard resolution slit of IFU1. The wavelengths from some of the bright  Hg lines are marked. Each order is labelled with its number, and the circles near the ends of the orders indicate the free spectral range.

The design of GHOST is optimised for $363 - 950$nm (orders $95 - 36$). The transition region around the beamsplitter is $523 - 544$\,nm, with these wavelengths appearing on both detectors (orders $66 - 64$). However, photons spanning the range $348 - 1061$\,nm fall on the detectors (orders $98 - 32$), with these wavelengths expected to be useful for science, at least for brighter targets. Regarding the spatial direction, a single object in standard resolution on the blue (red) camera is 27.8 (28.9) pixels long. In high resolution, a single object on the blue (red) camera is 45.3 (47) pixels long (measured at the centers of the camera).

Table~\ref{blaze} in the Appendix lists each order, its blaze wavelength $\lambda_B$, and its limiting wavelengths (corresponding to the wavelengths of the pixels at the edge of the detector, $\lambda_{min}, \lambda_{max}$). While these wavelength solutions are subject to small changes pending pressure variations and such like, we stress that GHOST data is effectively fixed format and very stable. The slight differences between the wavelength coverage of standard and high resolution is because the spatial separation of the high and standard resolution slits at the entrance to the bench spectrograph is in the wavelength direction when imaged on the science detectors. This results in a few angstroms difference in the wavelength limits at the edges of the detectors. 


\subsubsection{Quick look benefits of fixed format}

The fixed format of GHOST is especially convenient scientifically in terms of quick look capabilities. Indeed, the two object mode of GHOST makes possible quick-look studies of the relative properties of objects. For example, \cite{hayes2023} published the first scientific results from GHOST, obtained using commissioning data, describing the chemical abundance signature of a star identified in the Reticulum II dwarf galaxy, a nearby satellite of the Milky Way. The star had not previously been studied in detail, and its membership in the galaxy was unclear. The standard resolution mode of GHOST was used, with one of the IFUs positioned on the candidate member, and the second IFU positioned on a confirmed member star which was known to have an enhancement in elements formed via the rapid neutron capture process. 

\begin{figure}
   \begin{center}
   \includegraphics[width=8cm]{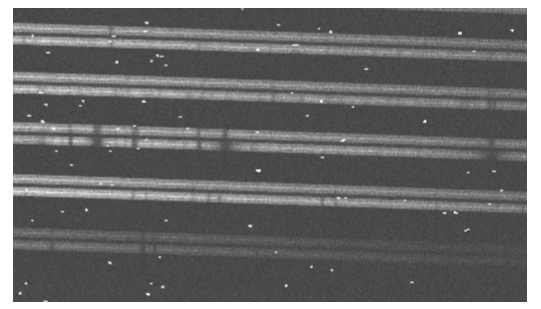}
   \end{center}
   \caption{ \label{ret2} A segment of the image from the blue camera from an observation of 2 stars in the Reticulum II dwarf galaxy, and analysed in detail in \cite{hayes2022}. The five pairs of spectra are five adjacent orders from IFU1 and IFU2 in the region surrounding the Mg b triplet. Bright spots are cosmic rays.}
\end{figure}

Figure~\ref{ret2} shows a segment of the image from the blue camera from this observation of Reticulum II stars. The five pairs of spectra are five adjacent orders from IFU1 and IFU2 in the region surrounding the Mg b triplet. A few things can be noted just from a visual inspection of the image. First, the image quality is very good: this is clear from the very bright central bands corresponding to a relatively bright central fiber compared to the fainter outer ring of fibers (see discussion in Section~\ref{sec:mapping}). Secondly, the two stars are clearly at almost the same velocity: this is clear from the fact that the absorption features are aligned. Given one of these stars is a confirmed velocity member of Reticulum II, this means the candidate is also a velocity member of the dwarf. Third, the star traced by the lower of the two spectra (the candidate) is considerably enhanced in Mg compared to the other, confirmed, member: this is clear from the much broader Mg b absorption line, visible in the central pair of spectra. More rigorous abundance analysis detailed in \cite{hayes2023} reveals that the candidate star is similarly enhanced in rapid neutron capture elements as the known member, but additionally has much higher light element abundances (for which Mg is a good tracer). This is a rare abundance pattern, and it is reasonable to question the methodology used in making the quantitative measurements. Here, being able to demonstrate that the methods produce results consistent with previous studies for the confirmed member, observed using the same set-up at the same moment as the unusual star, is a particularly good riposte.



 


\subsubsection{Delivered spectral resolving power and on-chip binning modes}

\begin{figure*}
   \begin{center}
   \includegraphics[width=\textwidth]{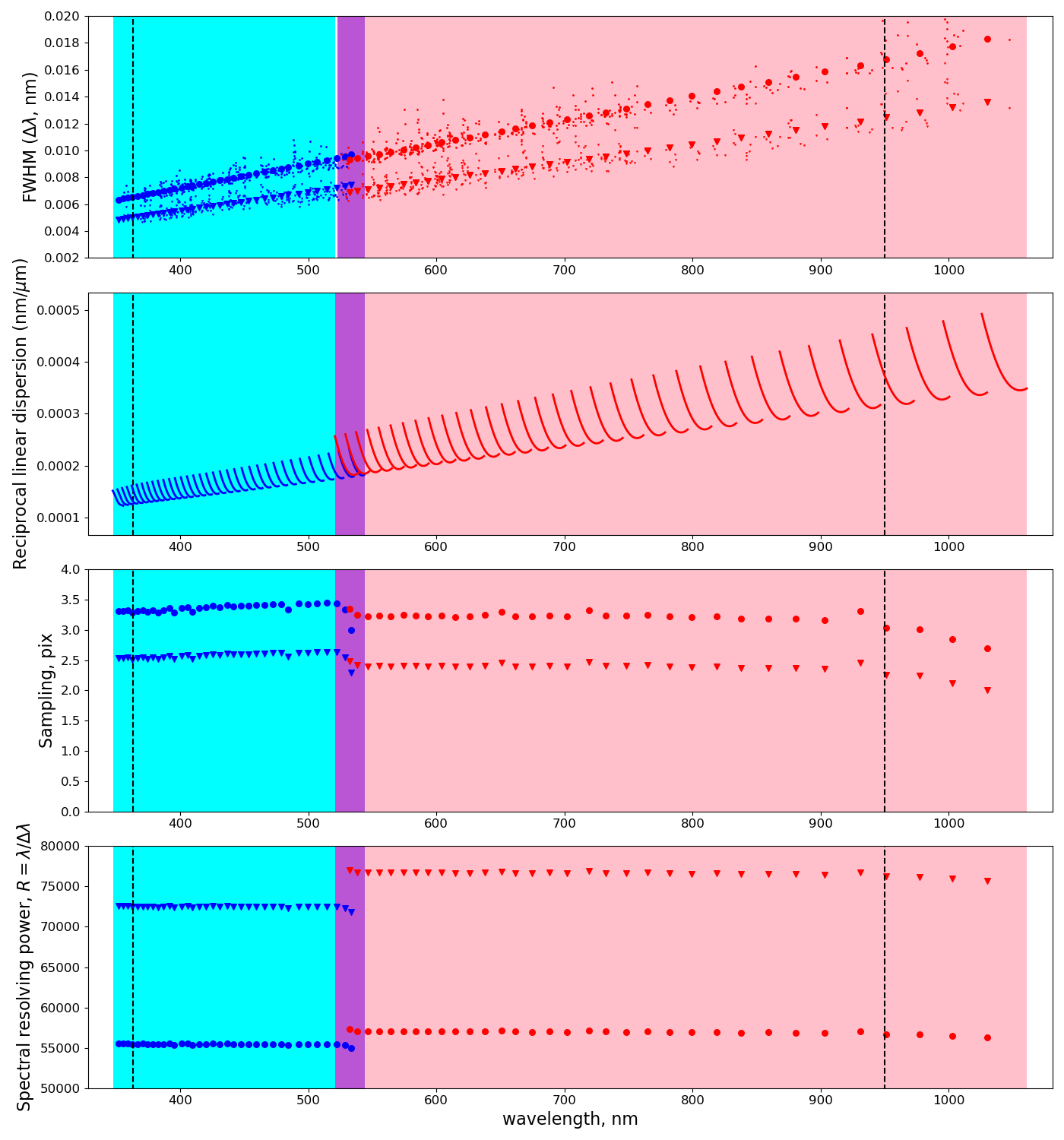}
   \end{center}
   \caption{ \label{specres} The top to bottom panels show, respectively, the spectral resolution ($\Delta\lambda$, the measured FWHM of unresolved lines), reciprocal linear dispersion, sampling per FWHM, and spectral resolving power ($R = \lambda/\Delta\lambda$) as a function of wavelength for the standard (dots) and high (triangles) resolution modes for the case of an evenly illuminated slit. The small points in the top panel show the raw measurements, and the larger points in the top, third, and bottom panels show the median values per order. See text for details.}
\end{figure*}

A variety of on-chip binning modes are available for GHOST, in both the spatial and spectral directions, and are summarised in Table~\ref{tab:binning}. Given that the slits are very nearly vertical everywhere, binning in the spatial direction leads to virtually no change in the spectral resolution, but can be highly beneficial for increasing the SNR of faint objects. Further, as shown later, GHOST is oversampled in the spectral direction in standard resolution mode, and so binning by 2 in the spectral direction in this mode does not significantly degrade the spectral resolution, while again greatly benefiting the SNR for fainter targets. In what follows, we describe the different spectral resolution and binning modes of GHOST and their recommended usage. We recognise that these recommendations are based largely on design consideration and future on-sky data for GHOST will better determine the utility of each mode for specific science cases.

\begin{table*}
\begin{center}
\begin{tabular*}{0.7\textwidth}{ccccp{3cm}}
IFU & Binning & Effective R&A\&G & Notes\\
 & spec $\times$ spat & && \\
\hline
  Standard & 1 x 1 & 56K & Direct (single or dual) &Oversampled; consider x2 binning in spectral direction\\ 
  & & & Companion guiding & \\
  & & & Blind offset & \\
  & & & Spiral search & \\
\hline
  Standard & 2 x 2 & 56K (46K) & Direct (single or dual) &Good default mode; marginally undersampled at 56K\\ 
  & & & Companion guiding & \\
  & & & Blind offset & \\
  & & & Spiral search & \\
\hline
  Standard & 2 x 4 & 56K (46K) & Direct (single or dual) &For faint objects\\ 
  & & & Companion guiding & \\
  & & & Blind offset & \\
  & & & Spiral search & \\
\hline
  Standard & 2 x 8 & 56K (46K) & Single direct &For faint objects\\ 
  & & & Blind offset & \\
  & & & Spiral search & \\
\hline
  Standard & 4 x 4 & (23K) & Direct (single of dual) &Spectral resolution set by detector sampling\\ 
  & & & Companion guiding & \\
  & & & Blind offset & \\
  & & & Spiral search & \\
\hline
  High & 1 x 1 & 75K & Single direct&Can be used with agitator, ThXe for PRV observations\\
  & & & Blind offset & \\
  & & & Spiral search & \\
\hline
  High & 1 x 2 & 75K & Single direct&Good default mode\\ 
  & & & Blind offset & \\
  & & & Spiral search & \\
\hline
  High & 1 x 4 & 75K & Single direct&For fainter objects\\ 
  & & & Blind offset & \\
  & & & Spiral search & \\
\hline\\
\end{tabular*}
\caption{Main configurations of GHOST in terms of delivered spectral resolving power for different on-chip binning options, as measured for even illumination of the slit. These should therefore be considered lower limits to values obtained when targeting on-sky point sources. Values in parentheses are those obtained assuming spectral resolving power is set by detector pixel sampling.}\label{tab:binning}
\end{center} 
\end{table*}

{\it Spatial binning considerations:} We first deal with spatial sampling considerations. For bright objects (i.e., where read noise is not a significant contribution to the noise budget), spatial binning by a factor of two will have limited impact given that the tilt of the slits in the spatial direction is very small (see Figure~\ref{rotate_slit}). However, the only case where spatial binning is certainly unadvisable are high resolution observations where the internal ThXe lamp is being used simultaneously with the science observation, which is anticipated to be anytime a precision radial velocity observation is being attempted. In this case, the gap between the ThXe spectrum and the main object spectrum on the detector is equivalent to only 1 high resolution microlens (see Figure~\ref{fibernumbers}). Binning by a factor of two therefore means that the ThXe and object spectrum will begin to merge.

Inspection of Figure~\ref{microlens_out} shows that the two object spectra of the standard resolution mode are separated by the sky fibers (the width of three microlens instead of the seven microlens that sets the length of an object spectrum). This equates to approximately 12 pixels on each camera separating the two objects in standard resolution mode.  Binning by 4 will therefore not merge the two objects, but binning by 8 will start to do so. Binning by 8 in the spatial direction is therefore not recommended if two objects are being targeted simultaneously. 

More generally, spatial binning - by a factor of 2, 4 or 8 in either mode - may be worthwhile depending on the luminosity of the targets and the contribution of sky, which itself is dependent upon the observing conditions. Specifically, spatial binning will merge the flux from the dedicated sky fibers in with the flux from the object(s). Even ignoring the dedicated sky fibers, there is a trade off between sky noise and read-noise that must be considered. As discussed in Section~\ref{sec:mapping}, the mapping of the object fibers onto the slit of GHOST is such that, in very good seeing, most of the object flux will be concentrated in the very center of the object spectrum, and most of the outer ``object'' fibers will be contributing significant amounts of sky. Given the slit camera images are used in the extraction of the spectrum, the data reduction process will account for this, and most of the outer object fibers will not contribute meaningfully to the extracted spectrum, meaning the excess sky in these fibers will not affect the measurement. If, on the other hand, spatial binning has been used, some of this sky flux will instead be combined with the object flux and the subtraction of this additional sky flux will introduce some noise into the extracted object spectrum.

{\it Spectral resolving power for unbinned observations} Figure~\ref{specres} shows the measured spectral resolution ($\Delta\lambda$), reciprocal linear dispersion, sampling and spectral resolving power ($R = \lambda/\Delta\lambda$) (top to botom panels, respectively) for standard (dots) and high resolution (triangles) modes as measured from ThAr (arc) calibration spectrum observed with the standard and high resolution modes, with $1 \times 1$ (spectral $\times$ spatial) binning. That is, this is the spectral resolving power of the spectrograph when the fibers are evenly illuminated, as would be the case in the limit of very poor seeing.

To create this figure, data were processed with the data reduction pipeline, including wavelength calibration. The reciprocal linear dispersion (nm per micron), displayed in the second panel, is easily measured from the wavelength solution by measuing the change in wavelength per pixel (recall this does not change between the standard and high resolution modes). We then fit Gaussian profiles to all of the strong lines to obtain a measure of the full-width-at-half-maximum (FWHM) of each line at every wavelength. We take the FWHM as a good approximation for the spectral resolution, $\Delta\lambda$). We inspected all fits and discarded any obviously poor fits, double lines, and other deviant measurements. Small points in the top panel of Figure~\ref{specres} show these direct measurements.

To reduce the scatter in the direct measurements, we estimate the median spectral resolution, sampling and spectral resolving power per order. First, for each order, we take the median FWHM per order. We assume that, on average, this represents the FWHM at the center (blaze wavelength) of each order i.e. all of these values across all orders represent the dispersion at the same diffraction angle from the grating. In this case, spectral resolution is then inversely proportional to order number i.e., FWHM $\propto 1/m$. We therefore fit this relationship to the median values in each camera for each mode of GHOST. This fit then gives us a robust estimate of the FWHM at the centers of each order, estimated from the data, but accounting for both measurement uncertainties and the fact that some orders only have a few lines (and thus a direct measurement of the median spectral resolution still has considerable uncertainty). These points are plotted as large symbols in the top panel, and are the only points shown in the third and fourth panels.

The spectral resolving power, sampling and reciprocal linear dispersion at the blaze wavelength in each order for these unbinned observations are given in Table~\ref{blaze} in the Appendix. At standard resolution, the median spectral resolving power in the blue (red) camera is 55.5K (57.0K), with samping of 3.4 (3.2) pixels, respectively.  At high resolution, the median spectral resolving power in the blue (red) camera is 72.5K (76.7K), with samping of 2.6 (2.4) pixels, respectively.  We emphasise that this is the spectral resolving power corresponding to even illumination of the slit, and is therefore a lower limit to the on-sky resolution when targeting a point source characterised by a seeing disk dependent upon the observing conditions.

{\it Spectral binning considerations:} Given the relative resolutions of the standard and high resolution modes, it is not envisioned that on-chip spectral binning in the high resolution mode will be necessary, when lower resolutions are achievable using the standard resolution mode instead. For the standard resolution mode, binning by a factor of 2 in the spectral direction reduces the sampling to 1.7 (1.6) pixels, which is only marginally undersampled and is expected to be sufficient and worth the trade-off for many science cases given the decrease in noise that the binning allows. As per the argument below, the effective spectral resolution given the mild undersampling may be expected to be closer to 45K.

On-chip binning in the spectral direction by a factor of 4 is also possible. In this case, the spectral resolution is not set by the PSF of the spectrograph optics or the size of the fibers, but instead is set by the detector pixel size. Using standard arguments, the spatial resolution of the detector can be considered equal to approximately two detector pixels  (although we note that this conceals many subtleties and we refer to reader to \cite{robertson2017} for a detailed discussion). An unbinned detector pixel is $15\mu$m, and so two binned detector pixels with $4 \times$ binning is equal to $120\mu$m. The reciprocal linear dispersion of the spectrograph (in units of nm per micron) is given in the second panel of Figure~\ref{specres}. The effective spectral resolution in the binned modes is therefore the product of the spatial resolution of the detector with the reciprocal linear dispersion, from which the spectral resolving power is wavelength divided by this quantity, in the usual way. This equates to a median resolving power of  23.4K in the blue arm, and 23.0K in the red arm. We note that this level of binning has not been used extensively thus far, but it is expected to be useful for observations of very faint stars for which the features under study do not require such high spectral resolving power as is achieved in the unbinned case. 






\subsection{Slit viewer camera}

\label{sec:slitcam}

GHOST has an Allied Vision G-238B slit viewer camera that images the output of the slit in blue and red light. Specifically, the beam from the spectrograph pseudo-slit enters the spectrograph via reflection from a custom beamsplitter. This beamsplitter transmits 1\% over all wavelengths, and the transmitted beam is imaged by a dedicated slit-viewing CCD camera in two wavelength bands (blue, 430 – 600 nm; red, 600 – 750 nm). The exposure time for this camera is set independently of the science exposures, and can be as short as 0.1 seconds. 

The slit-viewing camera has multiple uses. Possibly its most important use is during the data reduction process (\citealt{ireland2016, ireland2018, hayes2022} and Kalari et al., {\it preparation}; see also Section~5). The blue and red slit-viewing images closely correspond to the wavelength ranges of the blue and red science images, and they therefore provide an accurate measure of the flux distribution in the spatial direction for the optimal extraction of the science spectra. 

The slit images are also expected to be used as an exposure meter, and in order to calculate the photon-weighted mean exposure epoch for accurate correction for the Earth's motion. Indeed, even when the goal is not a precision radial velocity measurement, having a measurement of the integrated and time resolved distribution of light on the slit is useful for examing the (relative) observing conditions for any exposure, or between exposures. Indeed, since GHOST does not have an associated imager, the slit images are the only practical way to examine observing conditions (independent of Gemini telescope telemetry). For clear conditions with stable seeing, the total flux, and flux per fiber, in the slit camera images should be constant. Cloudy conditions are expected to affect the total flux from image to image, and variable seeing will change the flux distribution between fibers. The central fibers should generally always be the more illuminated given the distribution of flux across the IFU entrance: if this is consistently not the case, it is often a sign of, for example, the ADCs not being switched on or poor guiding.

\begin{figure*}
   \begin{center}
   \includegraphics[width=14cm]{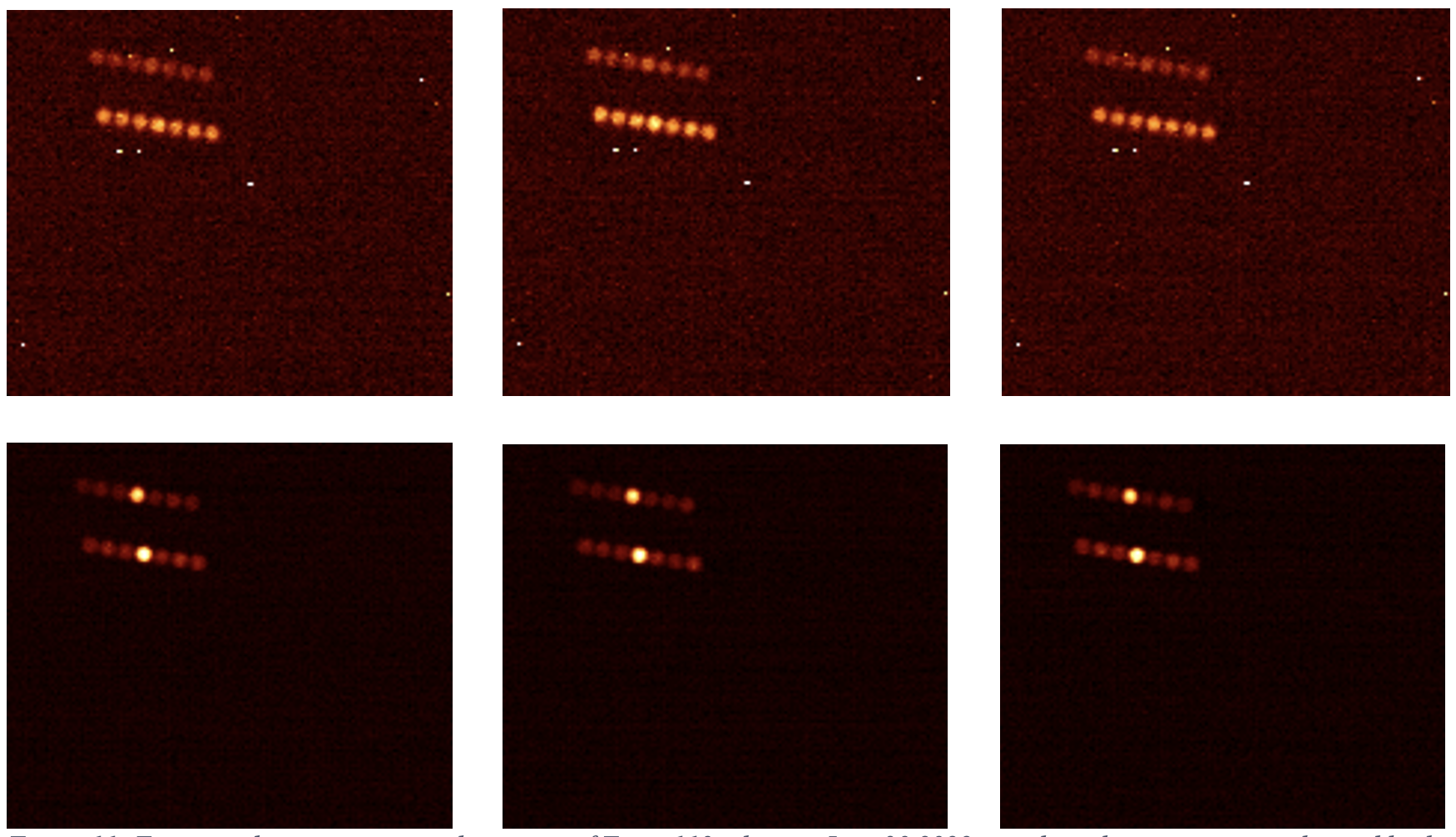}
   \end{center}
   \caption{ \label{svconditions} Top row: three consecutive slit images of Feige 110 taken on June 30 2022 in standard resolution, in relatively poor seeing and variable sky transparency. Bottom row: the same but for LTT\,3218 observed on January 29 2023. For the latter, the seeing is good and the sky transparency stable. For the slit images, the top image is the slit seen in red light, and the bottom image is the slit seen in blue light.}
\end{figure*}

\begin{figure*}
   \begin{center}
   \includegraphics[width=8cm]{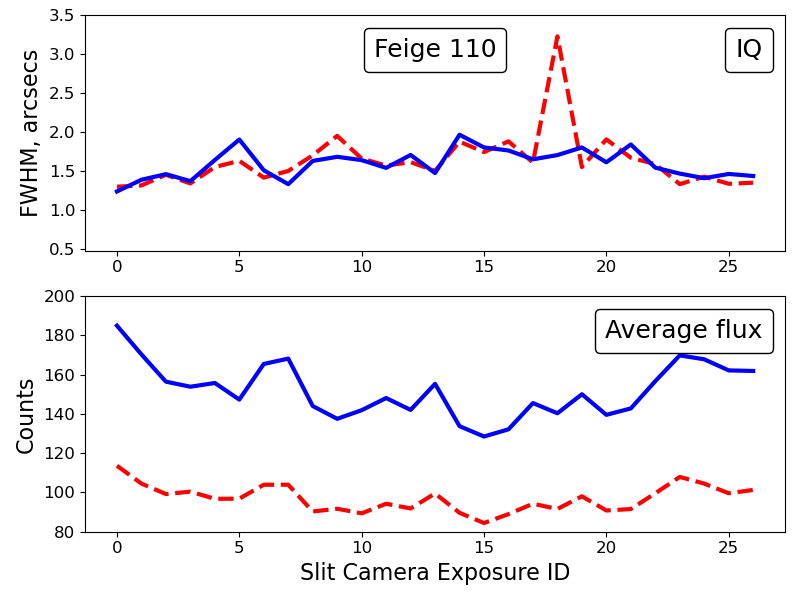}
   \includegraphics[width=8cm]{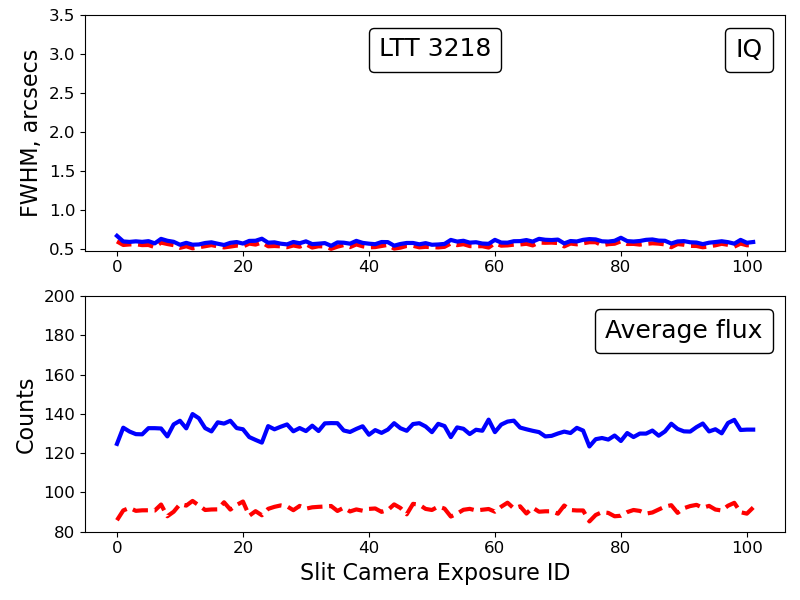}
   \end{center}
   \caption{ \label{iqflux} Top panels are estimates of the stellar FWHM based on consecutive slit camera exposures taken concurrently with the main science exposures. Bottom panels are the average counts per pixel in the illuminated IFU for each slit exposures. Calculations are made for both the blue and red slit images (blue and red lines, respectively). The left panels are for Feige 110, the right panels for LTT\,3218.}
\end{figure*}

As an example of the sort of diagnostics that the slit viewer camera enables, the top panel of Figure~\ref{svconditions} demonstrates the utility of the slit images in this regard. This shows three consecutive slit viewer images for an observation with poor seeing and unstable transparency  and variable conditions (June 28, Feige 110), while the bottom panels show the same but for good seeing and stable transparency (January 29, LTT\,3218). The two images of the slit in each panel correspond to the blue image (top) and red image (bottom) produced by the dichroic.

Quantitative analysis of the slit viewer images in Figure~\ref{svconditions} is revealing. Figure~\ref{iqflux} shows estimates of the seeing (top panels) and total flux (bottom panels) for these two stars (Feige 110 in the left panels, LTT\,3218 in the right panels). These are plotted sequentially for each slit camera image taken during the science exposures. IQ and flux are calculated independently for both the blue and red slit images. Note that the FWHM values assume a perfectly centered Gaussian PSF, and so are likely reasonable but not exact estimates of the actual FWHM. Nevertheless, it is clear that the flux in the images is stable for LTT\,3218, and the seeing is good and stable. For Feige 110, the flux is much more variable across the images, the seeing is worse, and more variable.



\subsection{Acquisition and guiding}

\label{sec:ag}

GHOST has its own acquisition and guiding system, independent of and not to be confused with the Gemini telescope acquisition and guiding system. Most GHOST observations begin by pointing the telescope at the necessary position on the sky, where it then begins guiding as normal using Gemini's peripheral wavefront sensors. Meanwhile, one or both GHOST IFUs are moved to their required position within the field of regard.

\begin{figure}
   \begin{center}
   \includegraphics[width=4cm]{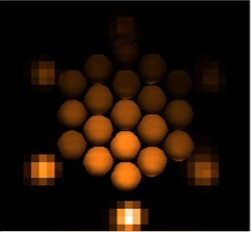}
   \includegraphics[width=4cm]{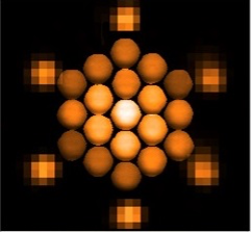}
   \end{center}
   \caption{ \label{acquistion} Left panel: reconstructed image of the fiber bundle immediately after acquisition by Gemini and Gemini guiding is started. Note the light distribution is not necessarily centered, given the pointing accuracy of Gemini. Right panel: Light distribution after a few iterations of the GHOST guiding, which seeks to balance the light distribution in the guide fibers.}
\end{figure}

\subsubsection{Direct acquisition}

\label{sec:direct}

The pointing accuracy of Gemini is of order $\sim 1$\,arcsec. In order to ensure that targets are more precisely centered in each IFU, each of the standard and high resolution science bundles is surrounded by a set of six guide fibers. These are shown as orange hexagons in Figure~\ref{microlens_in}. Unlike the science and sky fibers, these fibers are routed to an acquisition and guiding camera located next to GHOST in the pier lab of Gemini. The output images of this camera are analysed in real time, and an algorithm is applied that seeks to balance the flux in all six fibers, with small movements applied automatically to the relevant IFU positioners in order to achieve this and then maintain it during the course of the observation. Thus, a reconstructed image of the IFU using the guide camera and slit camera images immediately after Gemini guiding begins can be expected to look something like the left panel of Figure~\ref{acquistion}. Once GHOST guiding is switched on, the flux in each of the guide fibers is measured, and the algorithm better balances the flux in each of these fibers. After 4 or 5 guide camera exposures, the star is centered in the IFU, as shown in the right panel of Figure~\ref{acquistion}. Exposure times for this camera are set independently from the science camera, and can be as short as 0.1s, meaning the whole centering process initially takes less than a second for brighter stars. 

It is also possible for the guide camera to take a background sky exposure by moving the IFU to an appropriate region of sky. Once moved back to the target position, the measured guide fiber fluxes are then corrected for these sky values, and in this way direct acquisition and guiding is possible for fainter targets. For example, during commissioning, it was notable that direct acquisition was demonstrated to work successfully on a target with a magnitude of $G = 19.7$. Here, however, the guide camera images were 30 seconds in duration, and the initial centering necessarily took a few minutes.

One caveat of the direct acquisition procedure is that it is necessarily predicated on an azimuthally symmetric distribution of flux (i.e., a point source). Centering on an extended source, or indeed any source where the surroundings are not azimuthally symmetric, will require blind acquisition or companion guiding

\subsubsection{Blind acquisition offsets and companion guiding}

Alternatives exist for acquisition and guiding on faint targets or extended targets, where direct acquisition is not possible or too inefficient. The first of these, for acquisition only, is a blind offset. As the name suggests, this involves initially acquiring and centering on a nearby brighter star (within a few arcminutes). Once centered, an offset of the positioner in RA and declination ensures the target (a faint or extended source) is centered in the IFU. In this technique, the source starts the observation centered in the IFU because of the corrections that were applied on the offset star, but no active GHOST guiding is used during the observation.

The second technique, that is relevant only for guiding and only in the standard resolution mode, is ``companion guiding''. This is useful either to maintain active guiding on a faint star or extended source, or to maintain active guiding in a relatively crowded field, even on a relatively bright star. Here, the two IFUs are initially moved to their default positions to observe two targets. One of these targets must be a point source bright enough for direct acquisition and guiding. The other can be on an arbitrarily faint source, or in a relatively crowded field, or an extended source. GHOST acquisition and guiding is initiated as before but the positional corrections that are applied to both IFUs are derived using only the brighter target. 

Companion guiding was successfully demonstrated on-sky during commissioning. It is the only plausible way to actively guide on a star that is in the vicinity (few arcseconds) of a much brighter source. This is a known failure mode of the direct acquisition and guiding procedure. In this very specific situation, the guiding algorithm will naturally cause the IFU to eventually center upon the bright source if the outskirts of the bright source are of comparable luminosity to the wings of the target star that the guide fibers are using for centering.

\subsubsection{Spiral search}

The final acquisition method built into GHOST is a spiral search. This is anticipated to be used in rare circumstances where the coordinates of the target are not well known, or are off by a few arcseconds. Here, the source must be bright enough to be detected in a single acquisition and guiding image as per Section~\ref{sec:direct}. The IFU moves outwards in a spiral from a start position, taking an acquisiton and guide image and a slit camera image. If the total flux in the guide image is above a certain user-defined threshold, then GHOST concludes that the object has been found, and the procedure described in Section~\ref{sec:direct} to center on the target from this new starting position can be initiated.

\subsection{Precision radial velocity considerations}

\label{sec:prv}

\begin{table}
\begin{center}
\begin{tabular*}{0.6\columnwidth}{r|cc}
&\multicolumn{2}{c}{Max exposure time (s)}\\	
  ND filter & Blue & Red\\
 \hline\\
clear&1&1\\
0.5&1&1\\
1&1&1\\
1.5&5&1\\
2&15&3\\
2.5&90&10\\
3&150&15\\
3.5&1000&45\\
4&2500&120\\
5&$>3600$&300\\
\hline\\
\end{tabular*}
\caption{Exposure times for the ThXe lamp when used in conjunction with the neutral density filters, to prevent over-saturation of the lines. These times are indicative only and will need to be refined over the lifetime of GHOST.}\label{tab:thxe}
\end{center} 
\end{table}

Given the very benign physical environment of GHOST, the fixed spectral format, and the limited number of moving parts, some care has gone into enabling the possibility of it being used for precision radial velocities. Hardware includes a simultaneous calibration source (ThXe) for use in the high spectral resolution mode; fiber agitators for use when ultra-high SNR measurements are desirable; the slit camera, to track the slit illumination over the course of an exposure and to determine the photon-weighted mean exposure epoch.

The slit camera was discussed in Section~\ref{sec:slitcam}. The internal calibration source is only available for use with the high resolution mode. The source is very bright, and even a brief exposure can result in a reasonable number of saturated (Xe) lines in the red ($\gtrsim 8000$\AA). However, the lamp has its own selection of neutral density filters. Experimentation during commissioning suggested the combination of exposure times matched with filters summarised in Table~\ref{tab:thxe} is likely reasonable in that it results in a tolerable level of saturating, but this needs to be refined as more GHOST data is accumulated.

GHOST also has a fiber agitator to minimise modal noise for precision radial velocity measurements, and for any other observations requiring extremely high SNR measurements. In brief, optical fibers exhibit a non-uniform intensity distribution in the near and far fields due to the variability in mode excitation in the fiber. As a result, the changing states of interference between the modes results in a variable light output that is wavelength dependent. This intensity distribution may introduce wavelength shifts due to a variation of the response across individual pixels or adjacent pixels in the spectrograph. The GHOST fiber agitator physically displaces the optical cable over a fixed amplitude and frequency, acting as a mode scrambler. It more uniformly distributes the output optical power by rapidly changing the propagation lengths for all the propagating modes. Over the course of an observation, the intent is that this has the effect of making the near/far fields for a given wavelength more uniform, and thus maintains a more constant image for radial velocity measurements. The agitator can operate at a frequency between $0.5 - 1$\,Hz.

To excite the GHOST fibers, two voice coil positioning stages, mounted perpendicular to each other, are used to apply back and forth motion on the fibers. The voice coil stage is a compact actuator with a small footprint, and is ideal for closed-loop and short-stroke positioning applications that require accurate position, velocity, and acceleration control. It is mounted on linear bearings attached to the agitator mounting frame. The mounting frame is visible as the yellow structure in the photograph of GHOST in Figure~\ref{enclosure}, and the agitator itself is visible as the blue box on this yellow structure. We examine the delivered performance enabled by the agitators in Section~\ref{sec:modalnoise}.

\section{Science Performance}

\subsection{Ghosts in GHOST, and scattered light}

\begin{figure*}
   \begin{center}
   \includegraphics[width=8cm]{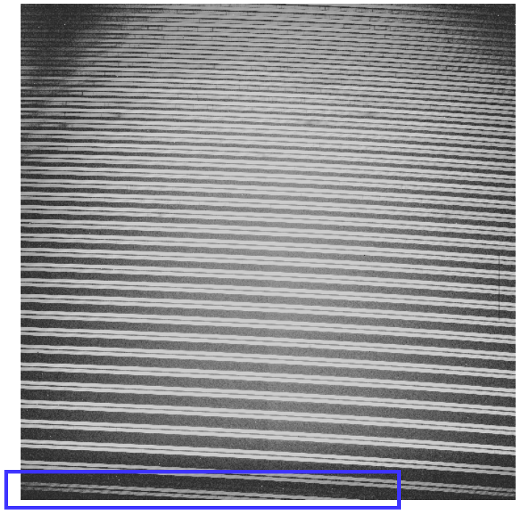}
   \includegraphics[width=8cm]{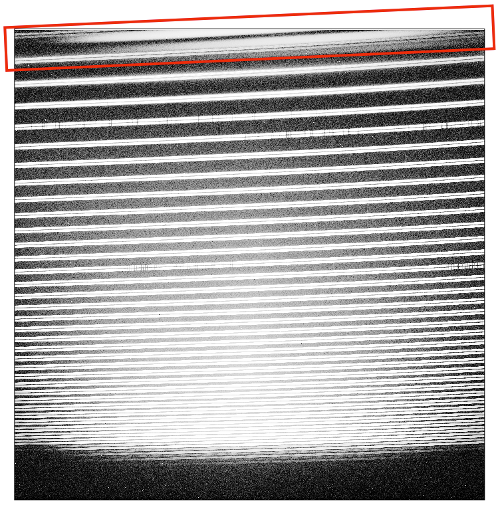}
   \includegraphics[width=8cm]{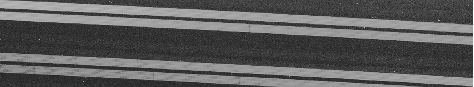}
   \end{center}
   \caption{ \label{ghostimage} Processed images from the blue (left panel) and red (right panel) detectors for an exposure targeting a pair of bright stars in standard resolution mode. The locations of all ghosts are highlighted, with a zoom-in of the blue ghost shown in the bottom panel.}
\end{figure*}

In what follows, we use the term {\it ghosts} to describe light that, while it is intended to be in the system, is not following the main optical path and which results in coherent, spurious, images on the detector. This is in contrast to scattered light, that does not produce coherent images but instead produces a diffuse background. Both of these are different than {\it stray light}, which refers to light that is not supposed to be in the system at all (and which is almost non-existent in GHOST - see Section~\ref{sec:detectors}). 

We performed checks for ghosting in GHOST during both integration at NRC and commissioning. During NRC integration, a few ghosts were identified on the blue camera when the instrument was illuminated with very blue light. For example, a  ghost is visible in Figure~\ref{hgspec} between orders 69 and 70. However, the illuminating source in this case is a Hg lamp, which is exceptionally bright in the blue and the UV, unlike any astronomical source being observed from the ground. Zemax modeling of GHOST, which had used a more realistic wavelength range, did not anticipate these ghosts.

In order to definitely check for ghosts relevant to science users of GHOST, we observed a pair of bright stars on the night of June 28 2022, HD122196 and HIP068460. These are two bright ($G = 8.51$ and $G = 8.58$, respectively), broadly equal luminosity stars (the first bluer, $G_{BP} - G_{RP} = -0.04$, the second redder, $G_{BP} - G_{RP} = 0.70$). We used standard resolution mode, positioning IFU1 on HD122196 and IFU2 on HIP068460. A single 900 second exposure was conducted in the blue, and a 750 second exposure was conducted in the red. The resulting images in the blue and red cameras both have peaks of around 30000 counts, with both IFUs illuminated. As such, this is a good test-bed to search for ghosts with the IFUs illuminated with realistic on-sky targets. 

The bias subtracted and mosaiced images from both detectors are shown in Figure~\ref{ghostimage}. For both images, we have applied various smoothing kernels and stretches to search for ghosts. None of the ghosts that showed up on the blue detector when using the blue lamp during integration and which are visible in Figure~\ref{hgspec} are present in these images, or on any image obtained during commissioning, in line with design expectations (including the Zemax models).

One ghost was discovered on each detector, however, and their locations are indicated by the  rectangle in Figure~\ref{ghostimage}. A zoom in and re-scaled image of blue ghost is shown in the bottom panel, and the faint ghost can just be discerned between the two main science orders (orders 63 and 64, counting from the bottom). The ghost overlaps slightly with order 63, which is a partial order, and does not cross order 64, which is the last full order on the blue detector. The counts in the ghost image are of order 4 counts/pixel above the background, whereas the peak counts in order 64 in this image are in excess of 900 counts/pixel.

For the red camera, a single set of ghost images is visible at the top of the detector using these commissioning data, and these ghosts were also previously identified during integration. They are highlighted with a box in the left panel of Figure~\ref{ghostimage}. These ghosts could potentially interfere with the extremely incomplete order $m = 31$ as well as the full order $m = 32$. Peak counts in the ghost image are approximately $4 - 5$\% of the peak counts in order 32. While this is not negligible, we note that order 32 corresponds to the extreme red of GHOST's wavelength range, which was originally anticipated to reach only $950$\,nm. While order 32 is visible on the detector, it is not extracted by default in the Data Reduction System due to generally low SNR for most targets. We expect the science impact of this ghost to be limited. We also note that, to our knowledge, no ghosts have been identified overlapping the main science orders of GHOST in any of the data that has been taken since commissioning, including during Science Verification.

In contrast to ghosts, that form coherent images on the detector, scattered light instead produces a diffuse background. The same images that were used to search for ghosts can also be used to reveal the scattered light levels. In both panels of the main panels of Figure~\ref{ghostimage}, scattered light can be seen between the main science orders, forming a non-uniform but slowly varying background. Its scale is modest: the peak of the scattered light in the blue is less than 1\% of the flux in the neighbouring orders. This is a typical value in the red detector as well, although here the peak of the scattered light relative to the flux in neighbouring orders reaches up to a few percent.

\subsection{Spectral stability and radial velocity accuracy}

\begin{figure*}
   \begin{center}
   \includegraphics[width=8cm]{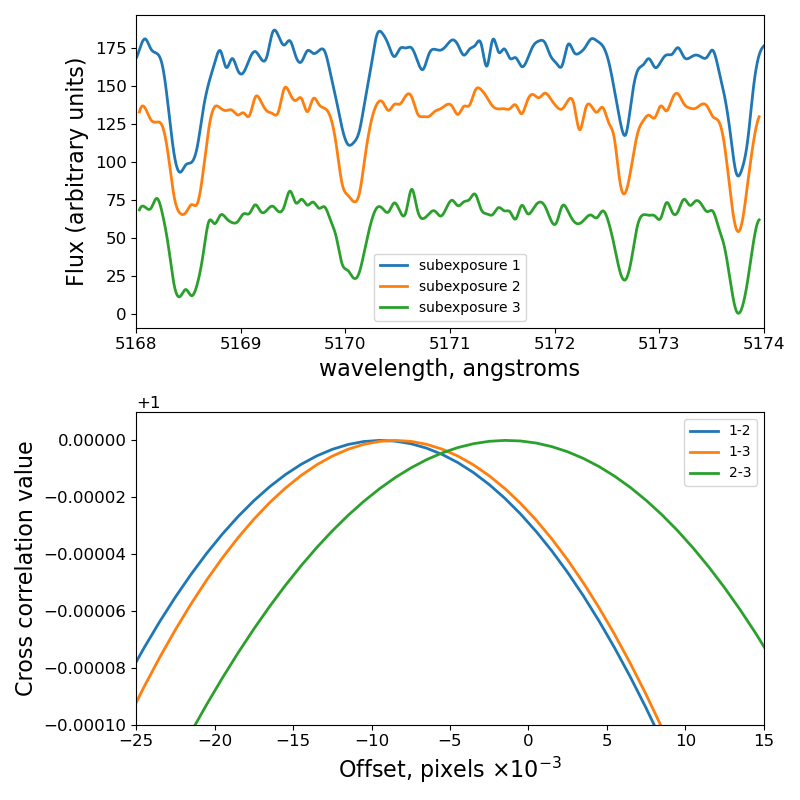}
   \includegraphics[width=8cm]{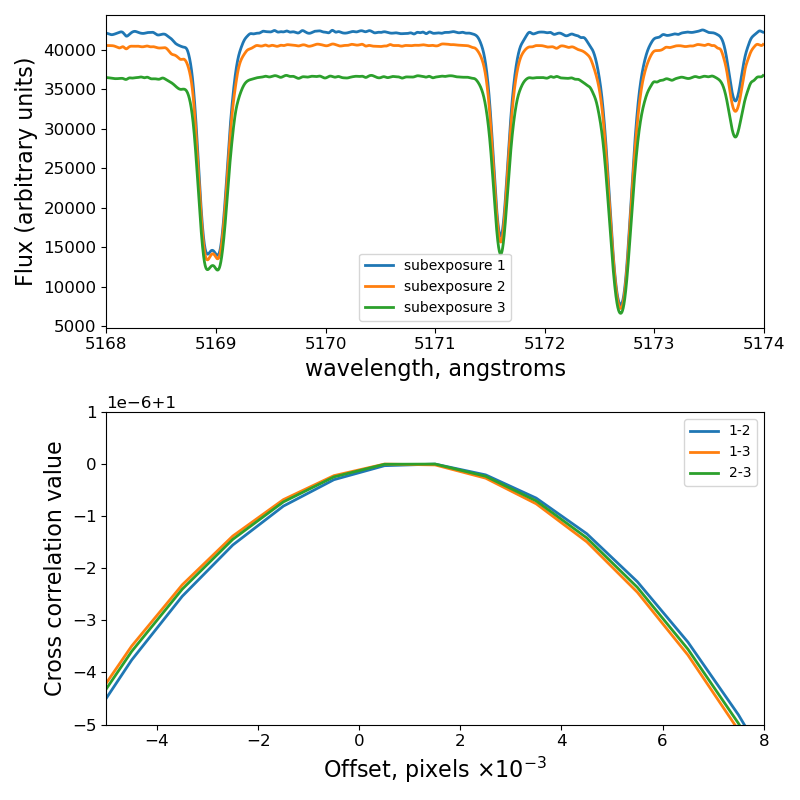}
   \end{center}
   \caption{ \label{stability} Top panels: individual subexposures for GDR-1760 (standard resolution, left) and HD122563 (high resolution, right) in the region around 5170\AA. Bottom panels: results of cross-correlation between the three subexposures.}
\end{figure*}

During the commissioning runs, most observations were taken as a set of repeat exposures, back-to-back (often sets of three exposures). This included several radial velocity standards that were also observed on multiple separate occasions (i.e. separated by a night or a few nights). Here, we examine the radial velocity precision as probed by the back-to-back data, meaning that we probe short duration timescales, and leave it to future work to demonstrate the longer-term velocity precision of GHOST.

For convenience, we use the standard and high resolution data published in \cite{hayes2023}. Details of these stars and their observations can be found in Tables 1 and 2 of \cite{hayes2023}. Details of the data reduction procedure for these stars can also be found in that paper.

While \cite{hayes2023} calculate radial velocities, we prefer to do a more basic check of the spectral stability than this, since their method of radial velocity determination also depends on the details of the code and templates they employ. Instead, we isolate a region of these spectra that has strong stellar features (enabling robust cross correlation), but which also appears to have limited contamination such as residual sky lines or telluric absorption features. Specifically, we use lines around 5170\AA, shown in the top panels of Figure~\ref{stability}, for GDR-1760 (standard resolution, left) and HD122563 (high resolution, right). In these panels, the three spectra correspond to the three sub-exposures, and have been arbitrarily offset in the y-axis for plotting purposes.  We interpolate the region 5168 – 5174\AA, with a cubic spline sampled every 0.01 pixels, and cross-correlate the three spectra with each other at each resolution. The cross-correlation peaks that are produced are shown in the lower panels of Figure~\ref{stability}, where the x-axis is in units of the original pixel, and the y-axis scale is arbitrary.

\cite{hayes2023} indicate in their Table 1 that, at this wavelength, the standard resolution spectra have a SNR of approximately 30. At this resolution, subexposure 1 and 2 are offset from each other by $\sim9.5$ resampled pixels; subexposure 1 and 3 are offset by $\sim8.5$ resampled pixels, and subexposures 2 and 3 are offset from each other by $\sim1.5$ resampled pixels. These offsets correspond to 305, 273, and 48 m/s, respectively.  At high resolution, the observations are at much higher SNR: \cite{hayes2023} indicate a SNR $\simeq 650$ (more than 20 times higher than the standard resolution data). Figure~\ref{stability} implies offsets between 0.5 and 1.5 resampled pixels (ie approximately 0.01 of a native pixel), or velocity offsets of 8 – 24 m/s.

This simple analysis suggests that, in standard resolution mode, for SNR $\simeq 30$ spectra, GHOST can provide radial velocity estimates precise to 300 m/s. For the high resolution mode, for SNR $\simeq 650$ spectra, velocity precision of 8 – 24m/s can be obtained. We recall that, fundamentally, the standard and high resolution modes of GHOST are opto-mechanically very similar and primarily differ only in slit width. Indeed, during GHOST Science Verification, we have been made aware of observations achieving velocity precision approaching 8m/s at standard resolution (Martioli et al., in preparation\footnote{see also \url{https://www.gemini.edu/sciops/instruments/gmos/SVobs/MartioliResults.pdf}}).  The formal requirement for GHOST velocity precision in this mode is 600m/s. Clearly, being able to reach a few hundred meters per second velocity precision on relatively faint stars is a science enabling capability, for example with respect to the internal velocity dispersion of ultra-faint dwarf galaxies, many of which remain unresolved using spectrographs with courser velocity resolution (e.g., see the compilation in \citealt{battaglia2022b}). At high resolution, the velocity requirement is 10m/s. We recall that precision velocity analyses can further take advantage of the simultaneous ThXe arcs to track shifts or changes in the wavelength scale, fiber agitators, the slit viewing camera to track changes in the input flux distribution,  and specialised data processing, none of which we have implemented in this analysis. The fact that, absent all of this, we are still approaching the required level of velocity precision at high resolution is a strong indication that GHOST should prove extremely useful for science in this velocity regime, although demonstrating the full precision radial velocity capabilities of GHOST at this early stage in its life is clearly beyond the scope of this current contribution. 

\subsection{Sensitivity}

\label{sec:sensitivity}

\begin{figure}
   \begin{center}
   \includegraphics[width=\columnwidth]{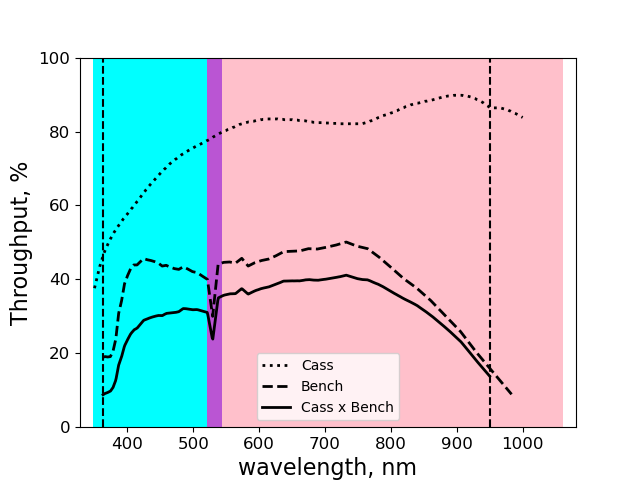}
   \end{center}
   \caption{ \label{throughput} The throughput of the Cassegrain unit, fibers and beam splitter (dotted line) and the throughput of the bench spectrograph (dashed line), including detectors, as they were measured during the integration phase. These combine to give the overall throughput of GHOST (excluding atmosphere, telescope and slit losses) shown by the solid line. The background colors are the same as in Figure~\ref{qe}.}
\end{figure}

The excellent throughput of some of the optical components of GHOST and the high detector QE were discussed in Section~\ref{sec:design}. These components, and others, combine to give the overall instrument throughput shown as the solid line in Figure~\ref{throughput}. This curve corresponds to the {\it measured} throughput of the GHOST instrument alone (i.e., not including the atmosphere, telescope, or slit losses. Specifically, the overall throughput of the Cassegrain unit, fibers and beam splitter were measured during integration of the subsystem (dotted line), and the overall throughput of the bench spectrograph (dashed line), including detectors, was measured during its integration. Figure~\ref{throughput} shows the result of combining these two empirical measurements, and demonstrates that the GHOST instrument from Cassegrain unit through to the bench spectrograph has a peak throughput of nearly 40\% at a wavelength of around 750nm.

What does this mean from the science user perspective? To address this critical question, multiple sets of spectrophotometric stars were observed during the GHOST commissioning runs. Given these are all spectrophotometric targets, they have a well-defined spectral energy distribution and  we know their AB magnitude at any wavelength in the optical. Examination of the slit viewer images that were obtained in parallel showed that many of these were unfortunately not taken in ideal conditions. However, the slit viewer images revealed those stars that were taken in the most stable conditions with the best IQ. Among them was LTT\,3218, previously shown in Figures~\ref{svconditions} and \ref{iqflux}, and taken with $2 \times 8$ binning (spectral $\times$ spatial), observed at an airmass of 1.00. For each of the observations identified as being taken in photometric conditions, we bias-subtract the data (but we do not apply a flat field), and we derive a wavelength solution. 

\begin{figure*}
   \begin{center}
   \includegraphics[width=\textwidth]{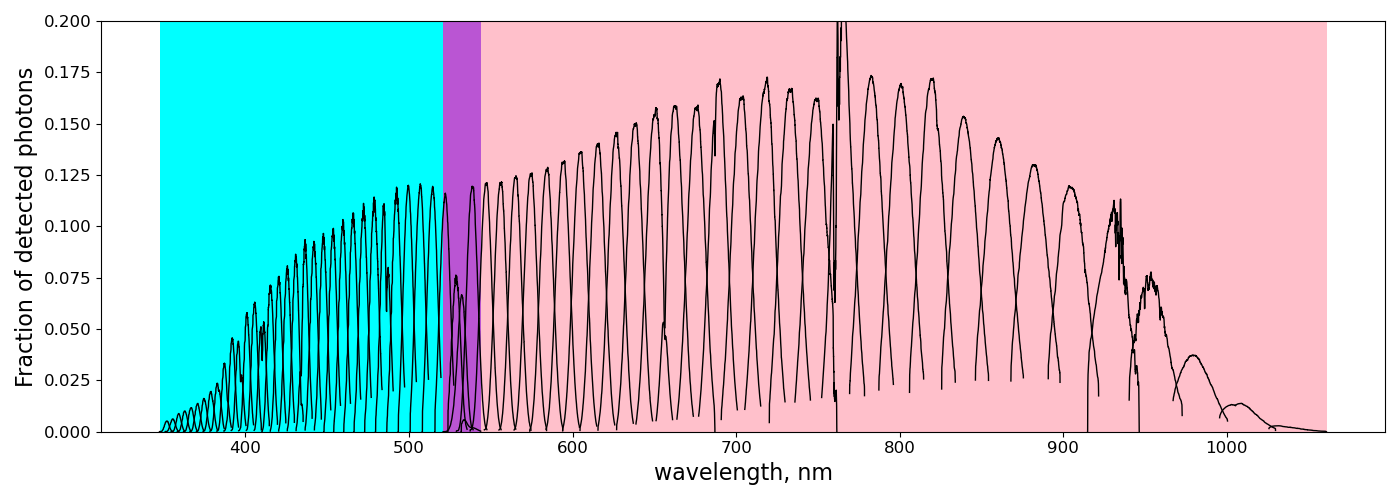}
   \includegraphics[width=\textwidth]{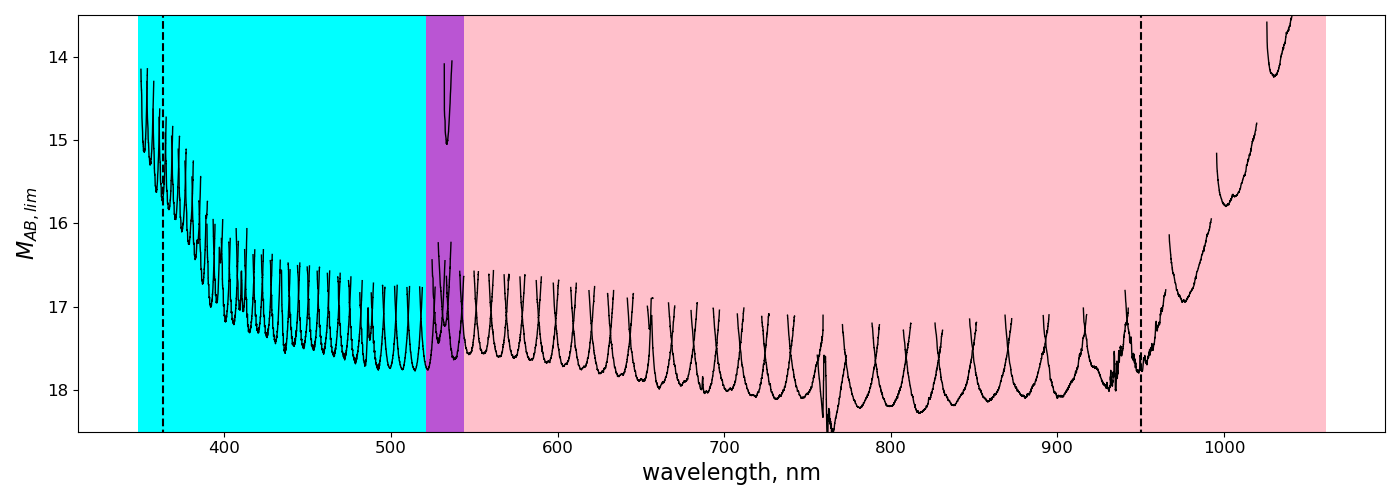}
   \end{center}
   \caption{ \label{limitingmag} Top panel: Fraction of photons detected by GHOST compared to those incident at the top of the atmosphere using on-sky observations of LTT\,3218, corresponding to dark skies, good seeing, an airmass of 1, and $2 \times 8$ detector binning.  Bottom panel: AB magnitude corresponding to a $SNR = 30$ per resolution element in a period of 1 hour, as a function of wavelength, for each order in GHOST at standard resolution, based on the measured efficiency derived in the top panel. For clarity, only wavelengths near the blaze wavelength of each order have been shown. Note, these sensitivity limits were calculated from observations taken with the Gemini primary mirror as it was in January 2023, some 7 years since its last recoating.}
\end{figure*}

For every pixel in every order of every subexposure, we are able to determine the number of electrons due to the star in the given exposure time. We can compare this to the number of photons incident at the top of the atmosphere from calibrated spectrophotometric measurements of this source. This overall efficiency includes atmospheric losses, slit losses, telescope losses, and spectrograph losses. The net throughput for an example observation of LTT\,3218 is shown in the top panel of Figure~\ref{limitingmag}. For wavelengths between approximately 450nm and 900nm, the overall efficiency of the Gemini/GHOST system exceeds 10\%, and peaks around 15\% between approximately 650nm and 800nm. We note that these observations were taken in January 2023, at a time when the Gemini primary mirror had not been recoated in some 7 years. We therefore expect the throughput of the overall system to be improved as a result of the recoating that occured in late 2023.

We can use this observation of LTT\,3218 to derive the sensitivity of GHOST in terms of the AB magnitude of the star that will produce a SNR per resolution element of 30 in a period of one hour. We can scale the number of counts that were received to the that which would be recorded from the star in a period of one hour. However, LTT\,3218 is very bright - Gaia $G = 11.8$ - and the observations of this star were relatively short, with an exposure time of 600s in the blue and 60s in the red. As such, we must additionally take into account read noise and sky noise, both of which will have a more significant contribution relative to source fluxes at faint magnitudes. Recall:


\begin{equation}
SNR \approx N_\star/\sqrt{(N_\star+ N_{sky} + n_{pix} \times e_r^2 )} \label{eq:snr}
\end{equation}

\noindent Here, $N_\star$ is the number of electrons due to the star, $N_{sky}$ is the number of electrons due to the sky, $n_{pix}$ is the number of pixels contributing to the measurement in the spatial direction, and $e_r$ is the read noise per pixel.  



Using Equation~\ref{eq:snr}, it is possible to solve for the required number of electrons per pixel that must be recorded in one hour to meet the SNR $= 30$ per resolution element requirement, under the observing conditions of LTT\,3218 and using the recorded number of sky counts as a function of wavelength. This can be compared to the number of counts received from LTT\,3218 given its known flux, to then derive the intrinsic limiting flux and corresponding limiting magnitude.

Systematic uncertainties will be introduced given that our measurement of the sky counts during the LTT\,3218 observation are very noisy given the very short exposures. We manage this latter issue by using a median filter over a moderate wavelength range to trace the sky continuum level, but we acknowledge the imperfection of this approach. Nevertheless, we prefer this empirical, on-sky, approach to demonstrate the overall sensitivity given the data available to us, compared to using synthetic sky counts.

Figure~\ref{limitingmag} shows the AB magnitude corresponding to a $SNR = 30$ per resolution element in a period of 1 hour, as a function of wavelength, for each order of each sub-exposure derived using the observation of LTT\,3218. For clarity, only wavelengths near the blaze wavelength of each order have been shown. Only the sensitivity at the blaze wavelength is immediately relevant given that the total sensitivity at wavelengths present in multiple orders requires summing over those orders.

We stress that the sensitivity limits described in Figure~\ref{limitingmag} and Table~\ref{blaze} take into account the full observing system, including the Gemini South telescope, and are strictly only correct for the conditions and setup under which LTT\,3218 was observed (dark skies, good seeing, airmass of 1, $2 \times 8$ detector binning). As previously emphasised, at the time of these observations, the Gemini primary mirror had not been coated for 7 years. The reflectivity of the primary mirror is known to be significantly degraded relative to a newly recoated mirror. As such, the sensitivity limits of GHOST on Gemini plan to be re-measured now that the Gemini South mirror has been recoated.


\subsection{Modal noise reduction at high SNR}

\label{sec:modalnoise}

\begin{figure}
   \begin{center}
   \includegraphics[width=8cm]{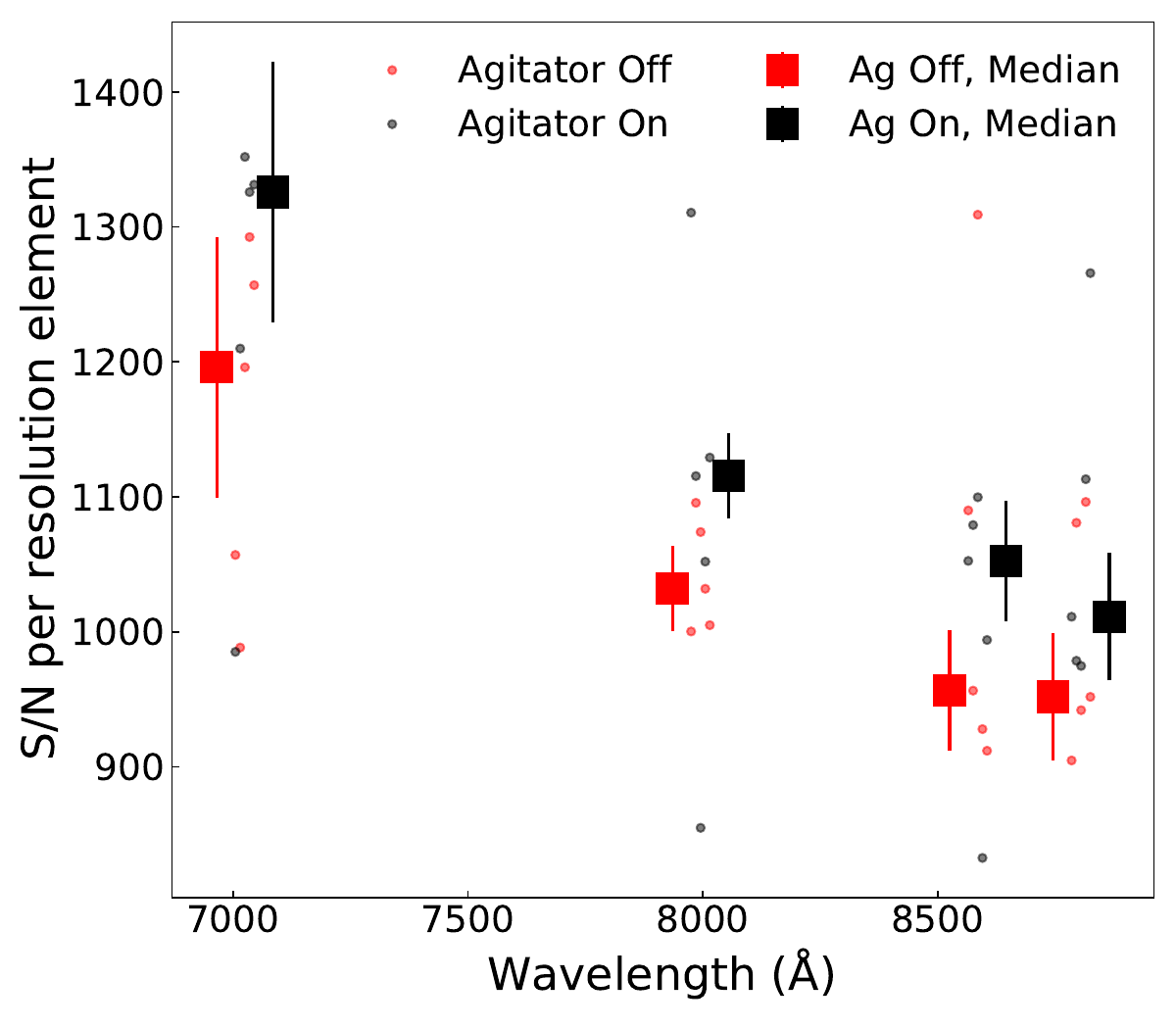}
   \end{center}
   \caption{ \label{modal} Signal-to-noise ratios per resolution element measured in quartz halogen flat field extracted spectra at four red camera order centers, taken with the GHOST fiber agitator off (red) and on (black).  Measurements were made in five segments of at least 200 pixels for each order and are shown as small points, and the median and median absolute deviation of these measurements are shown as large squares to better differentiate the effects of the fiber agitator on reducing modal noise.}
\end{figure}

As mentioned in Section \ref{sec:prv}, GHOST includes a fiber agitator that will move the fibers throughout an observation to average wavelength dependent interference patterns from different fiber propagation modes and mitigate the effects of fiber modal noise.  To illustrate the impact of modal noise in GHOST we compare the SNR ratios measured in flat lamp spectra with and without the fiber agitator active.

A sequence of 5x50s exposures of the GCAL 5W quartz halogen continuum lamp have been taken with an ND1 filter in the GHOST high resolution mode at two different telescope positions, Y$_1$ and Y$_2$, respectively, and with the fiber agitator on (set to an amplitude of 5mm and a frequency of 0.5 Hz) and off, Y$_{i, {\rm on}}$ and Y$_{i, {\rm off}}$, respectively.  These exposure times were chosen to be sufficiently short as to not saturate the red camera but long enough to provide a high SNR.

Each of these exposures were reduced and extracted using a 1D column extraction from the GHOSTDR pipeline, and those taken with the same telescope position and agitator status have been coadded to reduce photon noise.  The spectra at the two telescope positions have also been scaled to account for any lamp brightness variation between the two positions.  Using these spectra, we then take the difference and mean spectra between the flats taken at the two telescope position for the agitator on and off (e.g., D$_{\rm off}$ = Y$_{2, {\rm off}}$ $-$ Y$_{1, {\rm off}}$ and M$_{\rm off}$ = (Y$_{1, {\rm off}}$ $+$ Y$_{2, {\rm off}}$)/2).

The modal noise pattern changes with telescope position, so the difference spectrum with the agitator off has increased spectral variance due to modal noise, as compared to the difference spectrum with the agitator on.  Because the interference pattern from modal noise is coherent over the image of each fiber and the LSF in high resolution is approximately two pixels, the modal noise should result in correlations between neighbouring pixels in the difference spectrum with the agitator off.  To measure the effective pixel noise in the flat spectra we calculate the pixel noise including the neighboring pixel correlations:

\begin{equation}
\sigma_{\rm eff} = \sigma (1+ (cov)/ \sigma^2)
\end{equation}

\noindent Here, $\sigma^2$ and $cov$ (the covariance between adjacent pixels) have been calculated from the difference spectra and are divided by two to account for the fact that the difference spectra include two factors of the flat spectra noise.  The $\sigma_{\rm eff}$ noise per pixel is then converted to the noise per resolution element, $\sigma_{\rm pk} = \sigma_{\rm eff}/\sqrt{\Sigma B^2}$, where B is the normalised line spread function. 
Finally we calculate the SNR per resolution element of each flat by dividing the mean counts of the mean spectra by this noise per resolution element.

The number of modes in a fiber decreases with $\lambda^{-2}$ \citep{baudrand2001}, and therefore the effect of modal noise is expected to be largest at redder wavelengths, so we measure the SNR using the method described above in five bins of at least 200 pixels around the order centers of four, red GHOST orders (39, 40, 43, and 49 around the wavelengths of 8805, 8585, 7995, and $7025\,\AA$, respectively).  Figure \ref{modal} shows these SNR measurements, as well as the median and median absolute deviation of the measurements within each order, for both the agitator on and off.

While the individual measurements of the SNR have a relatively large scatter within each order, we do find, on average, a small impact of modal noise in GHOST spectra at or in excess of SNR $\sim 1000$, which can be mitigated through the use of GHOST's fiber agitator.  This implies that at SNR $\ll$ 1000 the fiber agitator is not needed, as photon noise will exceed the modal noise floor.

\section{Data reduction pipeline}

\label{sec:drp}

A complete summary of the data reduction processes for GHOST will be giving in Kalari et al. ({\it in preparation}; see also \citealt{ireland2018} and \citealt{hayes2022}). Briefly, GHOSTDR is the first Gemini instrument pipeline to be developed specifically for DRAGONS (Data Reduction for Astronomy from Gemini Observatory North and South; \citealt{labrie2023}), a python-based framework operating within the AstroConda environment. The core component of DRAGONS is the Recipe system, which is designed so that every image type (e.g., bias, dark, flat, arc, object) has an associated “recipe”, i.e., a list of processing steps referred to as “primitives”. The DRAGONS Reduce class automatically identifies the correct recipe for each image that is provided and processes it accordingly. This combined with the DRAGONS calibration database manager “caldb” will search for the appropriate calibration frames (biases, darks, flats, etc.) and supply these to the necessary primitives in the data reduction process. Combined, the recipe system and the calibration database mean that minimal user interaction is required to perform the basic data reduction steps.

GHOSTDR is an open source code built on the above framework, that is available at \url{https://github.com/GeminiDRSoftware/GHOSTDR}. While some primitives are standard in the DRAGONS system (e.g., bias and dark subtraction), GHOSTDR includes several primitives that are specific to the reduction of GHOST spectra. These include primitives such as order tracing and mapping of the integral field unit pseudo-slit onto the detectors, calculating 1D blaze corrections from flat fields, performing 2D extractions, fitting the wavelength solution, interpolating and order combining the extracted spectra, and optional capabilities such as determining the barycentric correction and response corrections with a spectrophotometric standard star.

For the order tracing, extraction and wavelength solution, GHOSTDR is built on a polynomial model of the spectrograph. This maps several quantities – the order trace, wavelength mapping, slit magnification in both spectral and spatial directions, and slit rotation – as a polynomial function of both order number and pixel number in the spectral direction. Due to the stability of the instrument, these models are expected to vary by only small amounts from night to night. Thus, default instrument models can be produced by an initial examination of high quality calibration frames and used as a starting point for the GHOSTDR reduction system to automatically perform fine tuned adjustments using nightly calibrations for science quality data reduction.

GHOSTDR fully incorporates its slit viewer camera which provides a direct measure of the slit profile to optimize the spectral extraction. Combined with the polynomial models of the spectrograph, this allow GHOSTDR to perform optimal extraction (\citealt{horne1986}) with a 2D extraction accounting for the the distribution of light across the slit and how the slit is imaged onto the detector (e.g., the slit rotation, etc.).

The pipeline offers multiple versions of the extraction, extracting objects from either IFU 1, IFU 2, or both, where any combination can include or ignore sky subtraction. For example, sky subtraction is necessary for deep observations of faint targets, but only adds noise for short exposures on bright targets. The reduction process generates intermediate data products along with their pixel variances and bad pixel maps. As a final data product, GHOSTDR returns two versions of the extracted 1D spectra for each camera: (i) order-by-order instrumental flux, and (ii) interpolated and order combined fluxes for each of the objects in the IFUs (and the sky IFU), along with the associated variances and wavelengths. 

\section{Summary}

GHOST was successfully commissioned at Gemini South in 2022. Since then it has been integrated into the Gemini instrumentation suite. The Observatory has conducted an extensive System Verification process during which time the end-to-end processes and hardware associated with GHOST, from proposal submission through to data reduction, were tested and de-bugged. Shared risk observing commenced in late 2023, with the instrument being available to the Gemini community from Semester 2024A onwards. It is the first new facilty-class instrument on Gemini for more than a decade, and it addresses a long-standing science need of the Gemini community. The first science results from GHOST, obtained using commissioning data, were presented in \cite{hayes2023}, recently followed by \cite{sestito2023c} and with more to come. Similarly, science results from Science Verification are also starting to appear (\citealt{placco2023, dovgal2024}).

GHOST is a late comer to a mature field. Its design takes advantages of advances in hardware and instrument design concepts to deliver a flexible work-horse instrument that its designers hope will enable new science and best-in-class performance.

\begin{acknowledgments}
The international Gemini Observatory, a program of NSF’s NOIRLab, is managed by the Association of Universities for Research in Astronomy (AURA) under a cooperative agreement with the National Science Foundation on behalf of the Gemini partnership: the National Science Foundation (United States), the National Research Council (Canada), Agencia Nacional de Investigaci{ó}n y Desarrollo (Chile), Ministerio de Ciencia, Tecnología e Innovaci{ó}n (Argentina), Minist{\'e}rio da Ci{ê}ncia, Tecnologia, Inova{ç}{õ}es e Comunica{ç}{õ}es (Brazil), and Korea Astronomy and Space Science Institute (Republic of Korea).

Supported by the international Gemini Observatory, a program of NSF’s NOIRLab, which is managed by the Association of Universities for Research in Astronomy (AURA) under a cooperative agreement with the National Science Foundation, on behalf of the Gemini partnership of Argentina, Brazil, Canada, Chile, the Republic of Korea, and the United States of America.
\end{acknowledgments}

\appendix

\section{Wavelength coverage and spectral resolving power for each order}

  \begin{center}
\begin{longtable}{ccc|ccccc|cccccc}
&&& \multicolumn{5}{c|}{High resolution mode}& \multicolumn{5}{c}{Standard resolution mode}\\
  $m$ & $\lambda_B$ & RLD & $\lambda_{min}$ & $\lambda_{max}$ &$\Delta\lambda$ & Sampling & $R$ &$\lambda_{min}$ & $\lambda_{max}$ &$\Delta\lambda$ & Sampling & $R$ \\
  & nm & nm/$\mu$m & nm & nm &nm & pixels per $\Delta\lambda$ & $\lambda/\Delta\lambda$ &nm & nm &nm & pixels per $\Delta\lambda$ & $\lambda/\Delta\lambda$\\
  \hline\\
  \endhead
  \multicolumn{3}{l|}{{\it Blue arm}}\\ 
98 & 351.9 & 0.00013 & 347.5 & 355.6 & 0.0049 & 2.53 & 72557 & 347.4 & 355.5 & 0.0063 & 3.31 & 55507\\
97 & 355.5 & 0.00013 & 351.1 & 359.3 & 0.0049 & 2.53 & 72551 & 351.0 & 359.2 & 0.0064 & 3.31 & 55502\\
96 & 359.3 & 0.00013 & 354.7 & 363.0 & 0.0050 & 2.54 & 72571 & 354.7 & 362.9 & 0.0065 & 3.32 & 55517\\
95 & 362.6 & 0.00013 & 358.5 & 366.8 & 0.0050 & 2.51 & 72474 & 358.4 & 366.7 & 0.0065 & 3.29 & 55443\\
94 & 366.7 & 0.00013 & 362.3 & 370.7 & 0.0051 & 2.54 & 72522 & 362.2 & 370.6 & 0.0066 & 3.31 & 55480\\
93 & 370.7 & 0.00013 & 366.2 & 374.7 & 0.0051 & 2.54 & 72533 & 366.1 & 374.6 & 0.0067 & 3.33 & 55489\\
92 & 374.4 & 0.00014 & 370.2 & 378.7 & 0.0052 & 2.53 & 72470 & 370.1 & 378.6 & 0.0068 & 3.30 & 55440\\
91 & 378.7 & 0.00014 & 374.2 & 382.9 & 0.0052 & 2.54 & 72505 & 374.1 & 382.8 & 0.0068 & 3.32 & 55467\\
90 & 382.5 & 0.00014 & 378.4 & 387.1 & 0.0053 & 2.52 & 72428 & 378.3 & 387.0 & 0.0069 & 3.29 & 55408\\
89 & 387.1 & 0.00014 & 382.7 & 391.5 & 0.0053 & 2.54 & 72485 & 382.6 & 391.4 & 0.0070 & 3.33 & 55451\\
88 & 391.9 & 0.00014 & 387.0 & 395.9 & 0.0054 & 2.58 & 72559 & 386.9 & 395.8 & 0.0071 & 3.37 & 55508\\
87 & 395.5 & 0.00014 & 391.5 & 400.5 & 0.0055 & 2.52 & 72393 & 391.4 & 400.3 & 0.0071 & 3.29 & 55381\\
86 & 400.9 & 0.00014 & 396.0 & 405.1 & 0.0055 & 2.58 & 72538 & 395.9 & 405.0 & 0.0072 & 3.37 & 55492\\
85 & 405.7 & 0.00014 & 400.7 & 409.9 & 0.0056 & 2.59 & 72553 & 400.6 & 409.7 & 0.0073 & 3.38 & 55504\\
84 & 409.5 & 0.00015 & 405.4 & 414.7 & 0.0057 & 2.52 & 72371 & 405.3 & 414.6 & 0.0074 & 3.29 & 55365\\
83 & 415.1 & 0.00015 & 410.3 & 419.7 & 0.0057 & 2.57 & 72487 & 410.2 & 419.6 & 0.0075 & 3.36 & 55454\\
82 & 420.2 & 0.00015 & 415.3 & 424.8 & 0.0058 & 2.58 & 72494 & 415.2 & 424.7 & 0.0076 & 3.37 & 55459\\
81 & 425.7 & 0.00015 & 420.5 & 430.0 & 0.0059 & 2.60 & 72547 & 420.3 & 429.9 & 0.0077 & 3.40 & 55499\\
80 & 430.7 & 0.00015 & 425.7 & 435.4 & 0.0059 & 2.58 & 72493 & 425.6 & 435.3 & 0.0078 & 3.38 & 55458\\
79 & 436.5 & 0.00015 & 431.1 & 440.9 & 0.0060 & 2.61 & 72551 & 431.0 & 440.8 & 0.0079 & 3.41 & 55502 \\
78 & 441.8 & 0.00016 & 436.6 & 446.5 & 0.0061 & 2.59 & 72502 & 436.5 & 446.4 & 0.0080 & 3.39 & 55465\\
77 & 447.6 & 0.00016 & 442.3 & 452.3 & 0.0062 & 2.60 & 72513 & 442.2 & 452.2 & 0.0081 & 3.40 & 55473\\
76 & 453.4 & 0.00016 & 448.1 & 458.3 & 0.0063 & 2.60 & 72498 & 448.0 & 458.1 & 0.0082 & 3.40 & 55462\\
75 & 459.6 & 0.00016 & 454.1 & 464.4 & 0.0063 & 2.61 & 72523 & 454.0 & 464.2 & 0.0083 & 3.41 & 55480\\
74 & 465.7 & 0.00016 & 460.2 & 470.6 & 0.0064 & 2.61 & 72505 & 460.1 & 470.5 & 0.0084 & 3.41 & 55467\\
73 & 472.2 & 0.00017 & 466.5 & 477.1 & 0.0065 & 2.62 & 72524 & 466.4 & 476.9 & 0.0085 & 3.42 & 55481\\
72 & 478.7 & 0.00017 & 473.0 & 483.7 & 0.0066 & 2.62 & 72515 & 472.9 & 483.5 & 0.0086 & 3.42 & 55475\\
71 & 484.3 & 0.00017 & 479.7 & 490.5 & 0.0067 & 2.55 & 72344 & 479.5 & 490.3 & 0.0088 & 3.34 & 55344\\
70 & 492.4 & 0.00017 & 486.5 & 497.5 & 0.0068 & 2.63 & 72518 & 486.4 & 497.3 & 0.0089 & 3.43 & 55477\\
69 & 499.4 & 0.00018 & 493.6 & 504.6 & 0.0069 & 2.62 & 72499 & 493.4 & 504.5 & 0.0090 & 3.43 & 55462\\
68 & 506.9 & 0.00018 & 500.8 & 512.1 & 0.0070 & 2.63 & 72521 & 500.7 & 511.9 & 0.0091 & 3.44 & 55479\\
67 & 514.5 & 0.00018 & 508.3 & 519.7 & 0.0071 & 2.64 & 72526 & 508.1 & 519.5 & 0.0093 & 3.45 & 55483\\
66 & 522.1 & 0.00018 & 516.0 & 527.5 & 0.0072 & 2.63 & 72499 & 515.8 & 527.4 & 0.0094 & 3.44 & 55462\\
65 & 528.7 & 0.00019 & 523.9 & 535.6 & 0.0073 & 2.55 & 72303 & 523.7 & 535.5 & 0.0096 & 3.33 & 55312\\
  64 & 533.6 & 0.00022 & 532.1 & 544.0 & 0.0074 & 2.29 & 71850 & 531.9 & 543.9 & 0.0097 & 3.00 & 54966\\
 
  \hline\\
  \multicolumn{3}{l|}{{\it Red arm}}\\ 
65 & 532.1 & 0.00019 & 521.0 & 539.4 & 0.0069 & 2.49 & 77047 & 520.9 & 539.2 & 0.0093 & 3.34 & 57281\\
64 & 538.5 & 0.00019 & 529.2 & 547.8 & 0.0070 & 2.41 & 76774 & 529.0 & 547.6 & 0.0094 & 3.25 & 57079\\
63 & 546.7 & 0.00020 & 537.6 & 556.5 & 0.0071 & 2.40 & 76726 & 537.4 & 556.3 & 0.0096 & 3.23 & 57042\\
62 & 555.6 & 0.00020 & 546.2 & 565.4 & 0.0072 & 2.41 & 76737 & 546.1 & 565.3 & 0.0097 & 3.24 & 57051\\
61 & 564.5 & 0.00020 & 555.2 & 574.7 & 0.0074 & 2.40 & 76709 & 555.0 & 574.5 & 0.0099 & 3.22 & 57030\\
60 & 574.2 & 0.00021 & 564.4 & 584.3 & 0.0075 & 2.41 & 76748 & 564.3 & 584.1 & 0.0101 & 3.24 & 57059\\
59 & 583.8 & 0.00021 & 574.0 & 594.2 & 0.0076 & 2.41 & 76730 & 573.8 & 594.0 & 0.0102 & 3.24 & 57046\\
58 & 593.6 & 0.00022 & 583.9 & 604.4 & 0.0077 & 2.40 & 76696 & 583.7 & 604.2 & 0.0104 & 3.22 & 57020 \\
57 & 604.1 & 0.00022 & 594.1 & 615.0 & 0.0079 & 2.40 & 76707 & 594.0 & 614.8 & 0.0106 & 3.23 & 57028\\
56 & 614.6 & 0.00022 & 604.7 & 625.9 & 0.0080 & 2.39 & 76671 & 604.6 & 625.8 & 0.0108 & 3.22 & 57002\\
55 & 625.8 & 0.00023 & 615.7 & 637.3 & 0.0082 & 2.39 & 76674 & 615.5 & 637.1 & 0.0110 & 3.22 & 57004\\
54 & 637.8 & 0.00023 & 627.1 & 649.1 & 0.0083 & 2.41 & 76724 & 626.9 & 648.9 & 0.0112 & 3.25 & 57041\\
53 & 650.9 & 0.00023 & 639.0 & 661.3 & 0.0085 & 2.46 & 76849 & 638.8 & 661.1 & 0.0114 & 3.30 & 57134\\
52 & 661.8 & 0.00024 & 651.2 & 674.0 & 0.0086 & 2.40 & 76662 & 651.0 & 673.8 & 0.0116 & 3.22 & 56995\\
51 & 674.7 & 0.00025 & 664.0 & 687.2 & 0.0088 & 2.39 & 76653 & 663.8 & 687.0 & 0.0118 & 3.22 & 56989\\
50 & 688.5 & 0.00025 & 677.3 & 700.9 & 0.0090 & 2.41 & 76687 & 677.1 & 700.7 & 0.0121 & 3.24 & 57014\\
49 & 702.1 & 0.00026 & 691.1 & 715.2 & 0.0092 & 2.39 & 76638 & 690.9 & 715.0 & 0.0123 & 3.22 & 56977\\
48 & 719.0 & 0.00025 & 705.5 & 730.1 & 0.0094 & 2.47 & 76881 & 705.3 & 729.9 & 0.0126 & 3.33 & 57158\\
47 & 732.3 & 0.00026 & 720.5 & 745.6 & 0.0096 & 2.41 & 76672 & 720.3 & 745.4 & 0.0128 & 3.24 & 57002\\
46 & 748.1 & 0.00027 & 736.2 & 761.8 & 0.0098 & 2.41 & 76660 & 735.9 & 761.6 & 0.0131 & 3.24 & 56993\\
45 & 765.0 & 0.00027 & 752.5 & 778.7 & 0.0100 & 2.42 & 76687 & 752.3 & 778.5 & 0.0134 & 3.25 & 57014\\
44 & 781.8 & 0.00028 & 769.6 & 796.4 & 0.0102 & 2.40 & 76630 & 769.4 & 796.1 & 0.0137 & 3.23 & 56971\\
43 & 799.5 & 0.00029 & 787.5 & 814.9 & 0.0104 & 2.38 & 76584 & 787.3 & 814.6 & 0.0140 & 3.21 & 56937\\
42 & 818.8 & 0.00030 & 806.2 & 834.2 & 0.0107 & 2.40 & 76609 & 806.0 & 834.0 & 0.0144 & 3.22 & 56955\\
41 & 837.9 & 0.00031 & 825.9 & 854.5 & 0.0109 & 2.37 & 76529 & 825.7 & 854.3 & 0.0147 & 3.19 & 56896\\
40 & 858.9 & 0.00032 & 846.5 & 875.8 & 0.0112 & 2.37 & 76534 & 846.3 & 875.6 & 0.0151 & 3.19 & 56900\\
39 & 880.7 & 0.00032 & 868.2 & 898.3 & 0.0115 & 2.37 & 76514 & 868.0 & 898.0 & 0.0155 & 3.19 & 56885\\
38 & 903.3 & 0.00033 & 891.1 & 921.9 & 0.0118 & 2.35 & 76466 & 890.8 & 921.6 & 0.0159 & 3.16 & 56849\\
37 & 931.0 & 0.00033 & 915.2 & 946.7 & 0.0121 & 2.46 & 76736 & 914.9 & 946.5 & 0.0163 & 3.31 & 57050\\
36 & 950.8 & 0.00037 & 940.6 & 973.0 & 0.0125 & 2.26 & 76250 & 940.3 & 972.7 & 0.0168 & 3.04 & 56689\\
35 & 977.3 & 0.00038 & 967.4 & 1000.7 & 0.0128 & 2.24 & 76198 & 967.1 & 1000.4 & 0.0173 & 3.01 & 56650\\
34 & 1002.8 & 0.00041 & 995.9 & 1030.0 & 0.0132 & 2.12 & 75953 & 995.6 & 1029.8 & 0.0178 & 2.85 & 56468\\
33 & 1030.0 & 0.00045 & 1026.0 & 1061.2 & 0.0136 & 2.00 & 75718 & 1025.7 & 1060.9 & 0.0183 & 2.69 & 56293\\
\caption{\label{blaze}For each order, $m$, this table lists the delivered blaze wavelength, $\lambda_B$, the reciprocal linear dispersion (RLD), the wavelength range, the sampling around $\lambda_B$, and the spectral resolving power around $\lambda_B$, for the high and standard spectral resolution modes. Note that order 32 also falls on the red detector, but it is extremely low SNR for most realistic observations and it is not currently extracted by the data reduction pipeline.}
\end{longtable}
\end{center}


%

\vspace{5mm}
\facilities{Gemini (GHOST)}


\software{DRAGONS (\citealt{labrie2023})}





\end{document}